\documentclass[reprint,superscriptaddress,aps,pra]{revtex4-2}

\usepackage{xcolor}
\usepackage{etoolbox} 
\usepackage{graphicx}
\usepackage{amsmath}
\usepackage{amssymb}
\usepackage{bm}
\usepackage[colorlinks=true,
            linkcolor=blue,
            citecolor=blue,
            urlcolor=blue]{hyperref}

\begin{document}

\title{Data-Efficient Learning of Anomalous Diffusion with Wavelet Representations: Enabling Direct Learning from Experimental Trajectories}

\author{Gongyi Wang}
\affiliation{School of Physics and Electronics, Hunan University, Changsha 410082, China}
\author{Yu Zhang}
\affiliation{School of Physics and Electronics, Hunan University, Changsha 410082, China}
\author{Zihan Huang}
 \email{huangzih@hnu.edu.cn}
\affiliation{School of Physics and Electronics, Hunan University, Changsha 410082, China}


\begin{abstract}
Machine learning (ML) has become a versatile tool for analyzing anomalous diffusion trajectories, yet most existing pipelines are trained on large collections of simulated data. In contrast, experimental trajectories, such as those from single-particle tracking (SPT), are typically scarce and may differ substantially from the idealized models used for simulation, leading to degradation or even breakdown of performance when ML methods are applied to real data. To address this mismatch, we introduce a wavelet-based representation of anomalous diffusion that enables data-efficient learning directly from experimental recordings. This representation is constructed by applying six complementary wavelet families to each trajectory and combining the resulting wavelet modulus scalograms. We first evaluate the wavelet representation on simulated trajectories from the andi-datasets benchmark, where it clearly outperforms both feature-based and trajectory-based methods with as few as 1000 training trajectories and still retains an advantage on large training sets. We then use this representation to learn directly from experimental SPT trajectories of fluorescent beads diffusing in F-actin networks, where the wavelet representation remains superior to existing alternatives for both diffusion-exponent regression and mesh-size classification. In particular, when predicting the diffusion exponents of experimental trajectories, a model trained on 1200 experimental tracks using the wavelet representation achieves significantly lower errors than state-of-the-art deep learning models trained purely on $10^6$ simulated trajectories. We associate this data efficiency with the emergence of distinct scale fingerprints disentangling underlying diffusion mechanisms in the wavelet spectra. These findings position the wavelet representation as a data-efficient basis for machine-learning analysis of anomalous diffusion, reducing reliance on simulation-based training and enabling more direct use of experimental trajectories.
\end{abstract}

\maketitle

\section{Introduction}\label{introduction}

Anomalous diffusion \cite{Metzler1,Klafter2005,Manzo2023preface,wang2012brownian}, a widespread phenomenon where transport deviates from classical Brownian motion, is prevalent across various complex systems such as biological tissues \cite{Illukkumbura2020,woringer2020anomalous,mine2017multi}, crowded environments \cite{skora2020macromolecular,regner2013anomalous,meyer2024time,Dechant2019,orre2021molecular,dai2022topology}, complex fluids \cite{mason1995optical,tavares2020anomalous,nakayama2024simulating,granick1991motions}, and financial markets \cite{Plerou2000,Masoliver2003,Jiang2019}. Effective analysis of anomalous diffusion is essential for revealing the dynamics and interactions within these systems. Beyond traditional statistical methods \cite{Sposini2022,Metzler2019,Vilk2022,Krapf2019,Burov2011,Thapa2018,Park2021,thapa2022bayesian,Krog2017,Krog}, machine learning (ML) techniques \cite{tan2019efficientnet,Dollar2021,dosovitskiy2020vit,ke2017lightgbm,chen2016xgboost,breiman2001random,vaswani2017attention,galton1886regression,cox1958regression,graves2012long,cho2014learning,korabel2025deep} have recently gained significant attention for their powerful ability to characterize anomalous diffusion trajectories, with applications including diffusion exponent prediction \cite{Munoz-Gil2,Li2021,Verdier2021,Verdier2022,Seckler2022,ahsini2025ai,haidari2025pointwise,Park2025}, diffusion model classification \cite{Kowalek2022,Loch2020,Granik2019,AL-hada2022,Firbas2023,DeepSPT2020,NOA2020,Pinholt2021}, and state transition detection (i.e., trajectory segmentation) \cite{Manzo2021,Bo2019,Argun2021,Dosset2016,Zhang2023,Wagner2017,Monnier2015,Matsuda2018,Gentili2021,Arts2019,Pineda2023,Qu2024,Requena2023,Martinez2023,seckler2024change,Chen2022,kabbech2022identification,Kabbech2024,bae2025segment,ahsini2025anomalousnet}. These ML methods have greatly enhanced the precision of anomalous diffusion analysis, and have the capacity to address challenges that conventional methods struggle to overcome.

Currently, the prevailing ML approaches for analyzing anomalous diffusion follow a {\em simulation-based} paradigm \cite{Munoz2021,munoz2025quantitative}, where models are trained on large synthetic datasets generated from idealized diffusion models. Once trained, these models are then transferred to real experimental data for analysis. By carefully designing simulated diffusion scenarios, the simulation-based paradigm leverages large amounts of labeled synthetic data, allowing ML models to capture complex patterns and relationships that would be difficult to discern using traditional methods \cite{cai2025machine,fernandez2024,asghar2024efficient,bender2024semore,saavedra2024supervised,cai1}. However, this paradigm relies heavily on the assumption that simulated data closely mirrors real experimental conditions, which is often not the case due to factors such as limited sample sizes, differences in diffusion processes, and distribution shifts between synthetic and real data \cite{ober2004localization,Simon2024,feng2024reliable}. As a result of this simulation-to-reality gap, models trained on simulated data may not generalize well to real-world scenarios, leading to performance degradation or even failure when applied to experimental data.

Despite these inherent limitations of the simulation-based paradigm when transferring to experimental data, it has so far remained the primary route for applying ML to anomalous diffusion. This situation is largely driven by practical constraints on experimental data. In typical experiments like single-particle tracking (SPT) \cite{Granik2019,sha2025single,wang2009anomalous,wang2010confining,chen2013diagnosing}, the total number of recorded trajectories is itself limited \cite{zhang25}. Moreover, since obtaining high-quality labels for these trajectories is substantially more difficult than generating labeled synthetic data, only a small fraction of experimental trajectories can be reliably annotated. Under such data constraints, the two main representation families for anomalous diffusion trajectories, namely feature-based \cite{Chhabra1989,Weron2017,Mangalam2023,Seckler2024,Pinholt2021} and trajectory-based \cite{Li2021,Granik2019,AL-hada2022,Firbas2023} representations, encounter specific limitations in the absence of simulation-based training, albeit for opposite reasons:

\begin{enumerate}

\item Feature-based representations map each trajectory to a vector of handcrafted descriptors with clear physical or statistical meaning, such as the mean squared displacement (MSD) and {\it p}-variation statistics. These physics-informed features allow models to remain reasonably reliable with limited experimental data, but their expressive power is constrained by the predefined feature set, so performance quickly saturates at a relatively low level even as more data are added \cite{Manzo2021}.
\item In contrast, trajectory-based representations operate directly on raw trajectories or step sequences, typically using deep sequence models to learn data-driven features. These high-capacity models can achieve excellent performance when trained on large datasets, but are extremely data-hungry and tend to severely underperform in the small-data regime (e.g., on the order of $10^3$ experimental trajectories) \cite{Li2021}.

\end{enumerate}

Taken together, these observations highlight that overcoming the limitations of the simulation-based paradigm cannot be achieved simply by accumulating more simulated data or scaling existing ML methods. Instead, relaxing this reliance on large synthetic datasets requires a trajectory representation of anomalous diffusion that combines physics-informed multiscale features with sufficient expressive capacity, enabling data-efficient and stable learning from limited experimental data while still fully exploiting large training sets when available.

In this work, we pursue such a representation by constructing a wavelet-based encoding of anomalous diffusion trajectories. By mapping trajectories to wavelet modulus scalograms on the time-scale plane, the characteristic multiscale structure of random motions is rendered explicit  and becomes more amenable to learning. On simulated trajectories from the andi-datasets benchmark \cite{Munoz2021}, the wavelet representation consistently outperforms both feature-based and trajectory-based approaches in diffusion-exponent regression and diffusion-model classification, even when trained on as few as 1000 trajectories. On experimental SPT trajectories of fluorescent beads diffusing in F-actin networks \cite{Granik2019}, models built on the wavelet representation remain superior to alternative representations for both diffusion-exponent regression and mesh-size classification. Notably, when predicting the experimental diffusion exponents, a model trained on 1200 experimental trajectories using the wavelet representation achieves substantially lower errors than state-of-the-art (SOTA) deep learning models following the simulation-based paradigm (trained purely on $10^6$ simulated trajectories). In addition, the characteristic scale fingerprints in the wavelet spectra are analyzed, which provide physical insights into the data-efficient behavior of the wavelet representation. These results demonstrate that the wavelet representation enables data-efficient learning of anomalous diffusion, thereby allowing direct learning from experimental trajectories and offering a promising direction toward developing alternatives to the simulation-based paradigm.

The rest of this paper is organized as follows. Sec. II presents the proposed wavelet representation in detail. Sec. III summarizes the simulated and experimental datasets, the associated learning tasks, and the evaluation metrics used in this work. Sec. IV reports the main results on simulated trajectories from the andi-datasets benchmark. Sec. V presents the results on experimental SPT trajectories of fluorescent beads in F-actin networks, specifically contrasting the proposed direct learning approach against the prevailing simulation-based paradigm. Sec. VI addresses interpretability by analyzing scale fingerprints in the wavelet spectra. Finally, we draw our conclusions in Sec. VII. Supporting information on wavelet scale ablation, experimental trajectory length distributions, detailed model-wise performance, noise optimization for simulation-based training, and wavelet family selection is provided in the Appendices.

\begin{figure}
\centering
\includegraphics[width=8.6cm]{Figure1.png}
\caption{Representative example of the continuous wavelet transform applied to a 1D trajectory. (a) Standardized 1D trajectory $x(t)$ of length $L=100$. (b) Corresponding wavelet modulus scalogram $|W_\psi(a, b)|$ computed with the real Morlet wavelet over 24 discrete scales.}
\label{fig:fig1}
\end{figure}

\begin{table*}
\caption{Continuous wavelet families used to construct the wavelet representation. Here, {\it Abbrev.} lists the shorthand names of wavelet families used throughout the text, {\it PyWT identifier} gives the corresponding callable name in PyWavelets, and {\it Parameters} specifies the wavelet parameters used in our implementation. {\rm j} denotes the imaginary unit.}
\label{tab:wavelets}
\begin{ruledtabular}
\renewcommand{\arraystretch}{2.3}
\begin{tabular}{lllll}
Continuous wavelet family & Abbrev. & PyWT identifier & Analytical form of $\psi(t)$ & Parameters \\
\hline
Real Morlet & morl &\texttt{morl} & $\psi(t)=e^{-t^{2}/2}\cos(5 t)$ &
-- \\
Mexican Hat & mexh& \texttt{mexh} &
$\psi(t)=\dfrac{2}{\sqrt{3}\,\pi^{1/4}}\,e^{-t^{2}/2}\,(1-t^{2})$ &
-- \\
Complex Gaussian Derivative& cgau& \texttt{cgau1} &
$\psi(t)\propto\dfrac{\rm d^P}{{\rm d}t^P}\left(\,e^{-{\rm j} t}\,e^{-t^{2}}\right)$ &
derivative order $P=1$ \\
Complex Morlet & cmor& \texttt{cmor1.5-1.0} &
$\psi(t)=\dfrac{1}{\sqrt{\pi B}}\,e^{-t^{2}/B}\,e^{{\rm j}2\pi C t}$ &
$B=1.5,\; C=1.0$ \\
Shannon & shan&\texttt{shan1.5-1.0} &
$\psi(t)=\sqrt{B}\,\dfrac{\sin(\pi B t)}{\pi B t}\,e^{{\rm j}2\pi C t}$ &
$B=1.5,\; C=1.0$ \\
Frequency B-Spline & fbsp&\texttt{fbsp3-0.5-0.5} &
$\psi(t)=\sqrt{B}\!\left[\dfrac{\sin\!\left(\pi B \tfrac{t}{M}\right)}{\pi B \tfrac{t}{M}}\right]^{M}
e^{{\rm j}2 \pi C t}$ &
$M=3,\; B=0.5,\; C=0.5$ \\
\end{tabular}
\end{ruledtabular}
\end{table*}

\section{Wavelet Representation}\label{wavelet}

In this section, we describe the proposed wavelet-based representation of anomalous diffusion in detail. For brevity, we refer to it as the {\em wavelet representation} throughout this paper. We start from a single standardized one-dimensional (1D) trajectory $x(t)$ of length $L$ and compute its continuous wavelet transform (CWT) \cite{grossmann1984decomposition,daubechies2002wavelet} with respect to a mother wavelet $\psi(t)$. The CWT is defined as:
\begin{equation}
W_\psi(a, b)=\int_{-\infty}^{\infty} x(t) \psi^*\left(\frac{t-b}{a}\right) \frac{\mathrm{d} t}{\sqrt{|a|}},
\end{equation}
where $a>0$ denotes the scale, $b$ is the translation (time position), and the superscript * denotes complex conjugation. For a fixed wavelet family, the complex coefficients $W_\psi(a, b)$ form a 2D time-scale representation of the trajectory, and we use their modulus $|W_\psi(a, b)|$ to construct a wavelet modulus scalogram. In this work, all CWTs are implemented using the Python package {\texttt {PyWavelets}} \cite{lee2019pywavelets}. A representative example is shown in Fig. \ref{fig:fig1}: a standardized 1D trajectory with $L=100$ [Fig. \ref{fig:fig1}(a)] is transformed into a modulus scalogram evaluated at 24 scales [Fig. \ref{fig:fig1}(b)] using the real Morlet wavelet \cite{torrence1998practical}. The color in this scalogram encodes the magnitude of $|W_\psi(a, b)|$, with localized high-amplitude patches revealing the multiscale structure of the underlying random motion.

\begin{figure*}
\centering
\includegraphics[width=17.0cm]{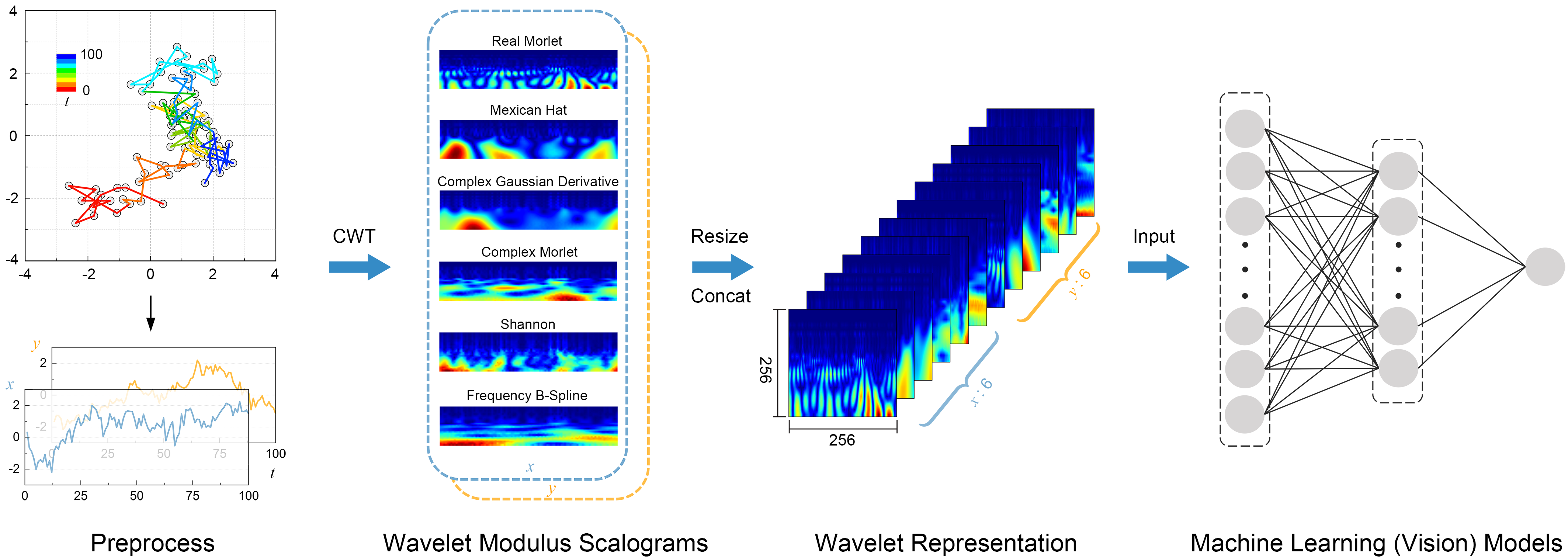}
\caption{Schematic construction of the wavelet representation. A multidimensional trajectory is first split into its 1D components, which are standardized. Subsequently, continuous wavelet transforms with six mother wavelets on the 24-scale grid of Eq.~\eqref{scaleset} produce wavelet modulus scalograms, which are resampled to a common $256\times256$ resolution and concatenated into a $6d$-channel tensor. This wavelet representation is then used as the input to downstream supervised (vision) models.}
\label{fig:fig2}
\end{figure*}

Six standard continuous mother wavelets are employed to construct the wavelet representation in this work:
\begin{enumerate}
\item Real Morlet wavelet \cite{torrence1998practical}: a sinusoidal carrier modulated by a Gaussian envelope, widely used for time-frequency analysis of oscillatory signals.

\item Mexican hat wavelet \cite{marr1980theory}: proportional to the second derivative of a Gaussian function, acting as a band-pass, pulse-like kernel sensitive to local peaks and curvature changes.

\item Complex Gaussian derivative wavelet \cite{addison2017illustrated}: a complex-valued wavelet defined as the $P$-th ($P=1$ in this work) derivative of a complex Gaussian function, providing simultaneous access to amplitude and phase information.

\item Complex Morlet wavelet \cite{teolis1998computational}: a Gaussian envelope multiplied by a complex sinusoidal carrier, with explicitly tunable bandwidth and center frequency.

\item Shannon wavelet \cite{cattani2008shannon}: a band-limited sinc function modulated by a complex exponential, corresponding to an idealized rectangular passband in the frequency domain.

\item Frequency B-Spline wavelet \cite{unser2002asymptotic}: a wavelet with a B-spline-shaped passband in the frequency domain, realizing a smooth, band-limited kernel with controllable main-lobe shape and smoothness.
\end{enumerate}
The detailed analytical forms of these mother wavelets, the abbreviations used throughout the text, and the corresponding PyWavelets identifiers and parameter settings are summarized in Table~\ref{tab:wavelets}.

Next, we adopt a fixed set of 24 scales for all CWT computations in this work:
\begin{equation}\label{scaleset}
\{0.1,0.2,\ldots,0.9,1,2,\ldots,15\},
\end{equation}
which spans from fine to coarse resolutions on the time-scale plane. For the wavelet families considered here, the pseudo-frequency associated with a given scale $a$ is approximately inversely proportional to $a$, so smaller scales tend to emphasize faster local fluctuations, whereas larger scales capture slower, more global structure. For discretely sampled trajectories with unit sampling interval, such a pseudo-frequency interpretation is only approximate, and scales $a<1$ may formally correspond to pseudo-frequencies beyond the Nyquist limit. However, since our goal is to construct an ML representation, we use the CWT primarily as a multiscale feature extractor rather than as a precise spectral estimator. To enrich the wavelet representation with short-time information, we explicitly include sub-unit scales $a \in \{0.1, 0.2, \ldots, 0.9\}$ to realize very short-support wavelet filters that are sensitive to variations over windows of one to a few time steps, such as abrupt changes in displacement or local fluctuations in step-to-step variability. On top of these fine scales, we use integer scales $a = 1, 2, \ldots, 15$ to capture slower variations in the overall spreading behavior and larger-scale trends along the trajectory. Preliminary experiments on simulated trajectories further indicate that extending the integer-scale range beyond $a = 15$ does not bring systematic improvement, whereas augmenting the integer scales with the sub-unit scales yields a consistent gain in downstream performance. We therefore adopt the combined set in Eq. \eqref{scaleset} as our default scale grid for all subsequent experiments. Detailed ablation results that support this choice are reported in Appendix \ref{scaleselection}.

After specifying the mother wavelets and the scale set in Eq.~\eqref{scaleset}, we now describe how the wavelet representation is constructed for a generic trajectory, as illustrated schematically in Fig. \ref{fig:fig2}. The detailed steps are as follows:
\begin{enumerate}
\item We consider a $d$-dimensional trajectory ${\bf x}(t) \in \mathbb{R}^d$ of length $L$ (with $d=1,2,3$ in this work) and decompose it into its $d$ 1D components. Each 1D component is then standardized independently by a per-trajectory normalization to zero mean and unit variance.
\item For each standardized 1D component, we compute the CWT with each of the six wavelet families in Table~\ref{tab:wavelets}, evaluated on the 24-scale grid in Eq.~\eqref{scaleset}. This yields, for each wavelet, a wavelet modulus scalogram $|W_\psi(a, b)|$ defined on a discrete time-scale grid of size $24\times L$. For a $d$-dimensional trajectory, this step produces $6d$ scalograms in total.
\item Each scalogram is subsequently resampled onto a common $256\times256$ grid in the time-scale plane. The resulting $6d$ scalograms are then concatenated along the channel dimension to form a $6d$-channel tensor, which we refer to as the wavelet representation of the trajectory.
\item This multi-channel representation serves as the input to downstream supervised models (in particular, computer vision models) for learning from the trajectories, without requiring any further task-specific feature engineering.
\end{enumerate}
This procedure yields a multiscale, image-like representation of trajectories that can be coupled with standard supervised models without additional feature engineering. In the following section, we detail the datasets and learning tasks used to assess its performance.

\begin{figure*}
\centering
\includegraphics[width=17.5cm]{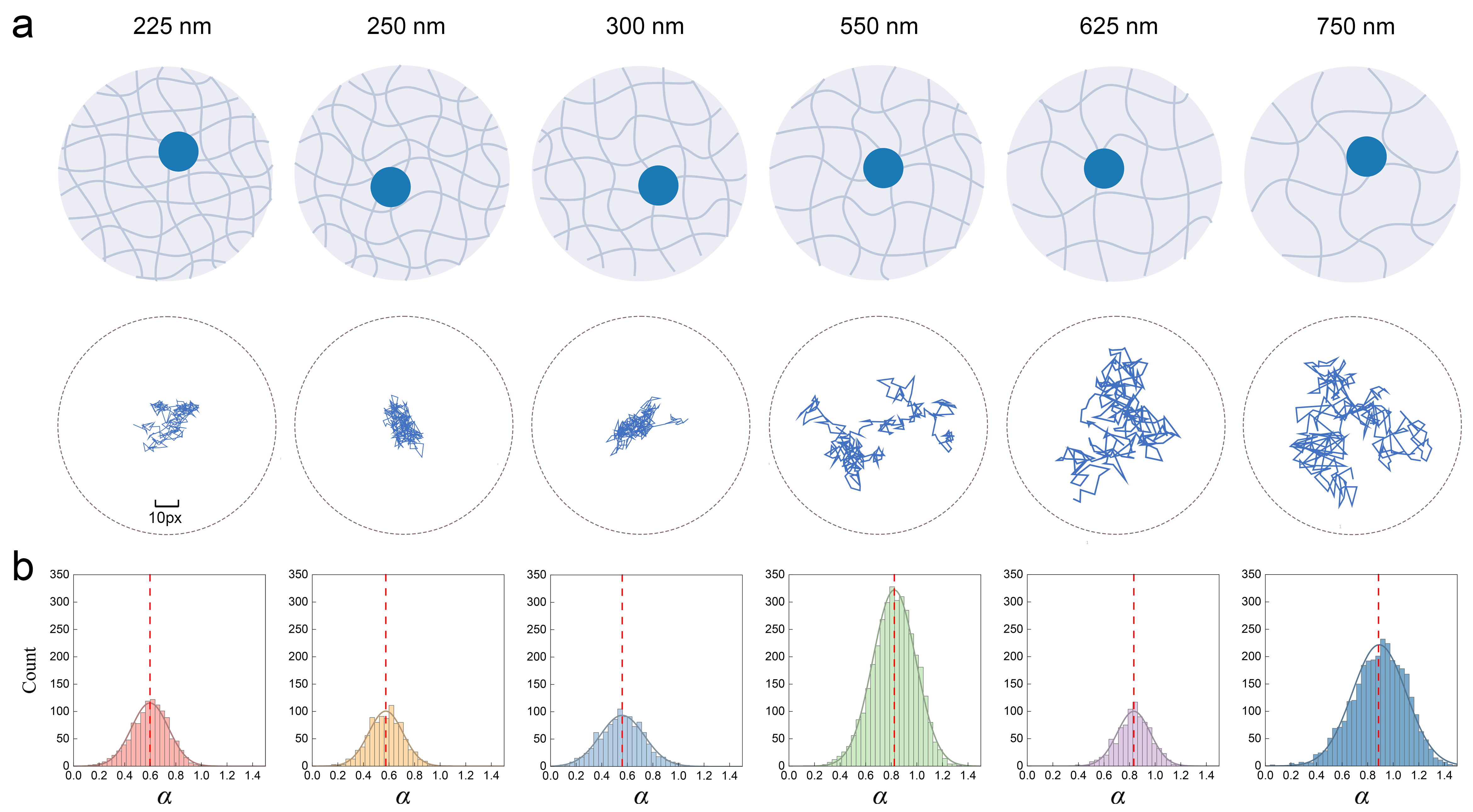}
\caption{(a) Schematic illustration of F-actin networks with mesh sizes $\xi \approx 225, 250, 300, 550, 625,$ and $750~\mathrm{nm}$ (top row), together with representative 2D bead trajectories recorded under each condition (bottom row). (b) Distributions of the diffusion exponent $\alpha$ estimated from TA-MSD fits for all trajectories at each mesh size; histograms show the counts of trajectories versus $\alpha$, and the red dashed line in each panel marks the mean exponent for that condition.}
\label{fig:fig3}
\end{figure*}

\section{Datasets and Tasks}

In this section, we present the anomalous diffusion datasets and learning tasks used to evaluate the proposed wavelet representation, together with the corresponding evaluation metrics. Sec. III A describes the simulated trajectories generated with the andi-datasets library \cite{Munoz2021} and the associated tasks of diffusion-exponent regression and diffusion-model classification. Sec. III B introduces the experimental SPT trajectories of fluorescent beads diffusing in F-actin networks with different mesh sizes \cite{Granik2019} and the corresponding task design.

\subsection{Simulated dataset and benchmark tasks}\label{task1}

The simulated anomalous diffusion trajectories used in this work are generated with the open-source \texttt{andi-datasets} library, originally developed for the AnDi Challenge and now widely used as a benchmark for testing analysis methods on anomalous diffusion data. This library implements five prototypical stochastic models of anomalous transport covering a broad range of diffusion exponents $\alpha$:
\begin{enumerate}
\item Annealed transient time motion (ATTM) \cite{Massignan}: a non-ergodic process in which the particle undergoes Brownian motion with a diffusivity that is randomly renewed in time or space, leading to effective exponents in the range $0.05 \leq \alpha \leq 1$.
\item Continuous-time random walk (CTRW) \cite{Scher}: a random-walk model with broadly distributed waiting times between successive steps, giving rise to subdiffusive dynamics with $0.05 \leq \alpha \leq 1$.
\item Fractional Brownian motion (FBM) \cite{Mandelbrot}: a Gaussian process driven by fractional Gaussian noise with power-law temporal correlations, which can generate both subdiffusive and superdiffusive behavior, $0.05 \leq \alpha < 2$.
\item L\'evy walk (LW) \cite{Klafter1}: a superdiffusive model in which step durations and jump lengths are coupled, producing non-Gaussian displacement statistics and exponents in the range $1 \leq \alpha \leq 2$.
\item Scaled Brownian motion (SBM) \cite{Lim}: a diffusion process with a deterministically time-dependent diffusion coefficient, yielding $0.05 \leq \alpha \leq 2$.
\end{enumerate}
Here, the diffusion exponent $\alpha$ characterizes the time scaling of MSD (${\rm MSD} \sim t^\alpha$). Simulated trajectories are generated from these five diffusion models in one, two, or three spatial dimensions with uniformly spaced time steps and optional additive Gaussian localization noise.

\begin{table*}
\caption{ML models used for different data representations. Here, {\it Representation} lists the type of input representation, {\it ML model} gives the model family, {\it Implementation details} specifies the hyperparameters or implementation used in our experiments, and {\it Abbrev.} lists the shorthand names of ML models used throughout the text.}
\label{tab:mlmodels}
\begin{ruledtabular}
\renewcommand{\arraystretch}{1.3}
\begin{tabular}{llll}
Representation & ML model & Implementation details & Abbrev. \\
\hline
Wavelet & EfficientNet \cite{tan2019efficientnet}&
\texttt{timm/efficientnet\_b5.sw\_in12k\_ft\_in1k} \cite{eff}& EffNet \\
 & RegNet \cite{Dollar2021}& \texttt{timm/regnetz\_040\_h.ra3\_in1k} \cite{reg}& RegNet \\
 & Vision transformer \cite{dosovitskiy2020vit}& \texttt{timm/vit\_huge\_patch14\_224.orig\_in21k} \cite{vit}& ViT \\
Feature & LightGBM \cite{ke2017lightgbm}& \texttt{n\_estimators=100, max\_depth=5} & LGB \\
 & XGBoost \cite{chen2016xgboost}& \texttt{n\_estimators=100, max\_depth=5} & XGB \\
 & Random forest \cite{breiman2001random}& \texttt{n\_estimators=100, max\_depth=None} & RF \\
 & Linear regression \cite{galton1886regression} & / & LinR \\
 & Logistic regression \cite{cox1958regression} & \texttt{C=1.0} & LogR \\
Trajectory & Transformer \cite{vaswani2017attention}& Ref. \cite{Requena2023} & Trm \\
 & WADNet \cite{Li2021}& Ref. \cite{wad}& WADNet \\
 & LSTM \cite{graves2012long}& 3 recurrent layers, hidden size 64 & LSTM \\
 & GRU \cite{cho2014learning}& 3 recurrent layers, hidden size 64 & GRU \\
\end{tabular}
\end{ruledtabular}
\end{table*}

On this simulated dataset, we consider two supervised learning tasks. The first is the {\em diffusion-exponent regression} task, where the goal is to infer the diffusion exponent $\alpha$ of a single trajectory. Performance on this regression task is quantified by the mean absolute error (MAE):
\begin{equation}\label{MAE}
\mathrm{MAE}=\frac{1}{N} \sum_{i=1}^N\left|\alpha_{i, \mathrm{p}}-\alpha_{i, \mathrm{GT}}\right|,
\end{equation}
where $\alpha_{i, \mathrm{p}}$ and $\alpha_{i, \mathrm{GT}}$ denote the predicted and ground-truth exponents of the $i$-th trajectory, and $N$ is the number of test trajectories. The second is the {\em diffusion-model classification} task, in which each trajectory must be assigned to one of the five diffusion models \{ATTM, CTRW, FBM, LW, SBM\}. As in the AnDi Challenge, we evaluate the classification performance using the micro-averaged F1 score:
\begin{equation}\label{F1}
\mathrm{F1}=2\cdot\dfrac{\text {precision} \cdot \text {recall}}{\text {precision}+ \text {recall}}.
\end{equation}
Here, precision and recall are computed by pooling true positives (TPs), false positives (FPs), and false negatives (FNs) over all five classes, and are given by:
\begin{equation}
{\text {precision}} = \frac{\rm TP}{\rm TP+FP},~{\text {recall}} = \frac{\rm TP}{\rm TP+FN}.
\end{equation}

\subsection{Experimental SPT dataset and task design}\label{task2}

\begin{figure*}
\centering
\includegraphics[width=17.5cm]{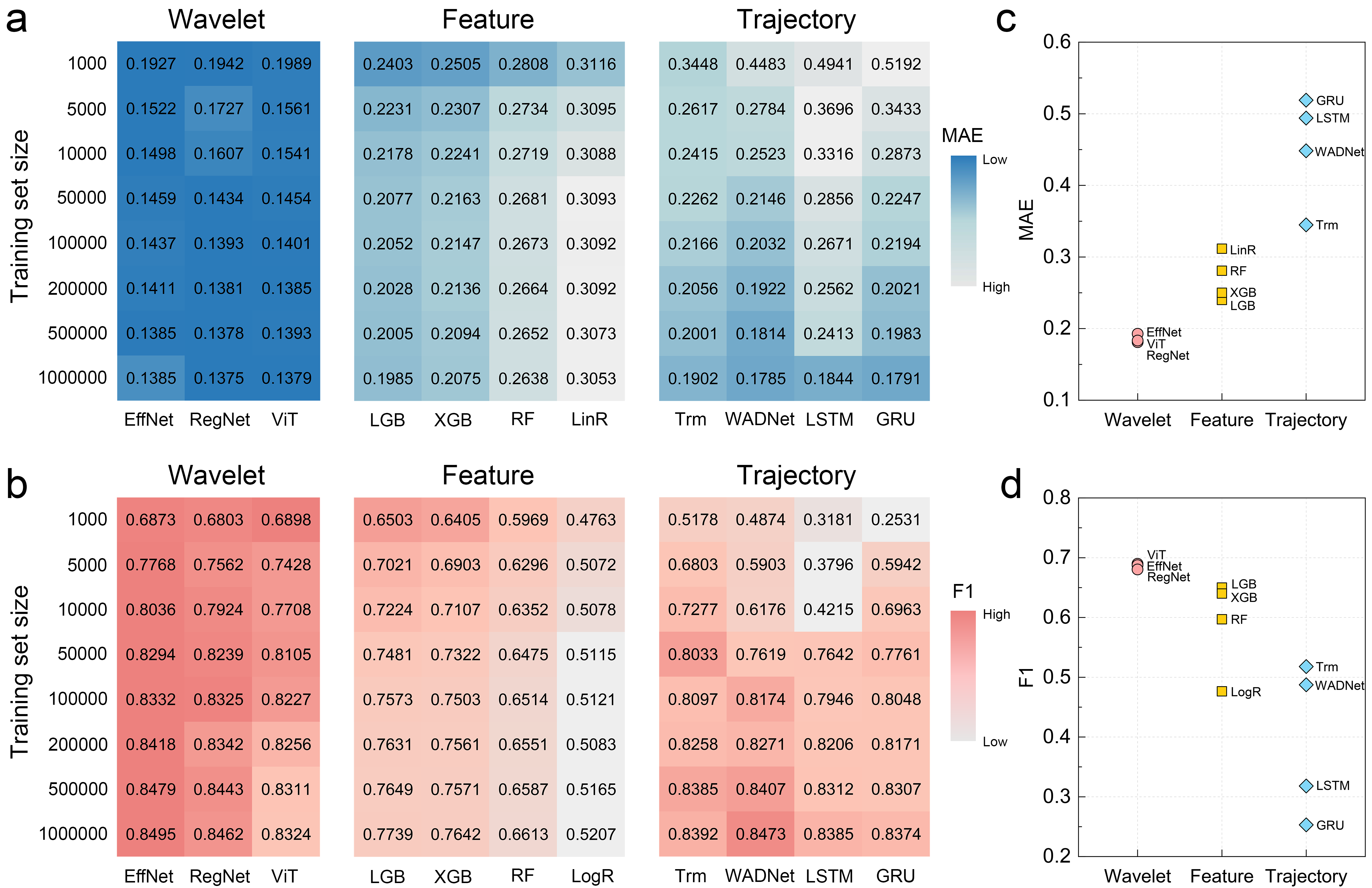}
\caption{Performance comparison of wavelet-, feature-, and trajectory-based representations on 2D simulated trajectories of length $L=100$. (a)-(b) Heatmaps displaying the performance on the test set across varying training set sizes $N_{\rm train}$. (a) MAE for diffusion-exponent regression (blue scale; darker indicates lower error). (b) Micro-averaged F1 scores for diffusion-model classification (red scale; darker indicates higher F1). (c)-(d) Detailed performance comparison at the small-data regime $N_{\rm train}=1000$, highlighting a clear advantage of wavelet-based models over both feature-based and trajectory-based alternatives.}
\label{fig:fig4}
\end{figure*}

\begin{figure*}
\centering
\includegraphics[width=17.5cm]{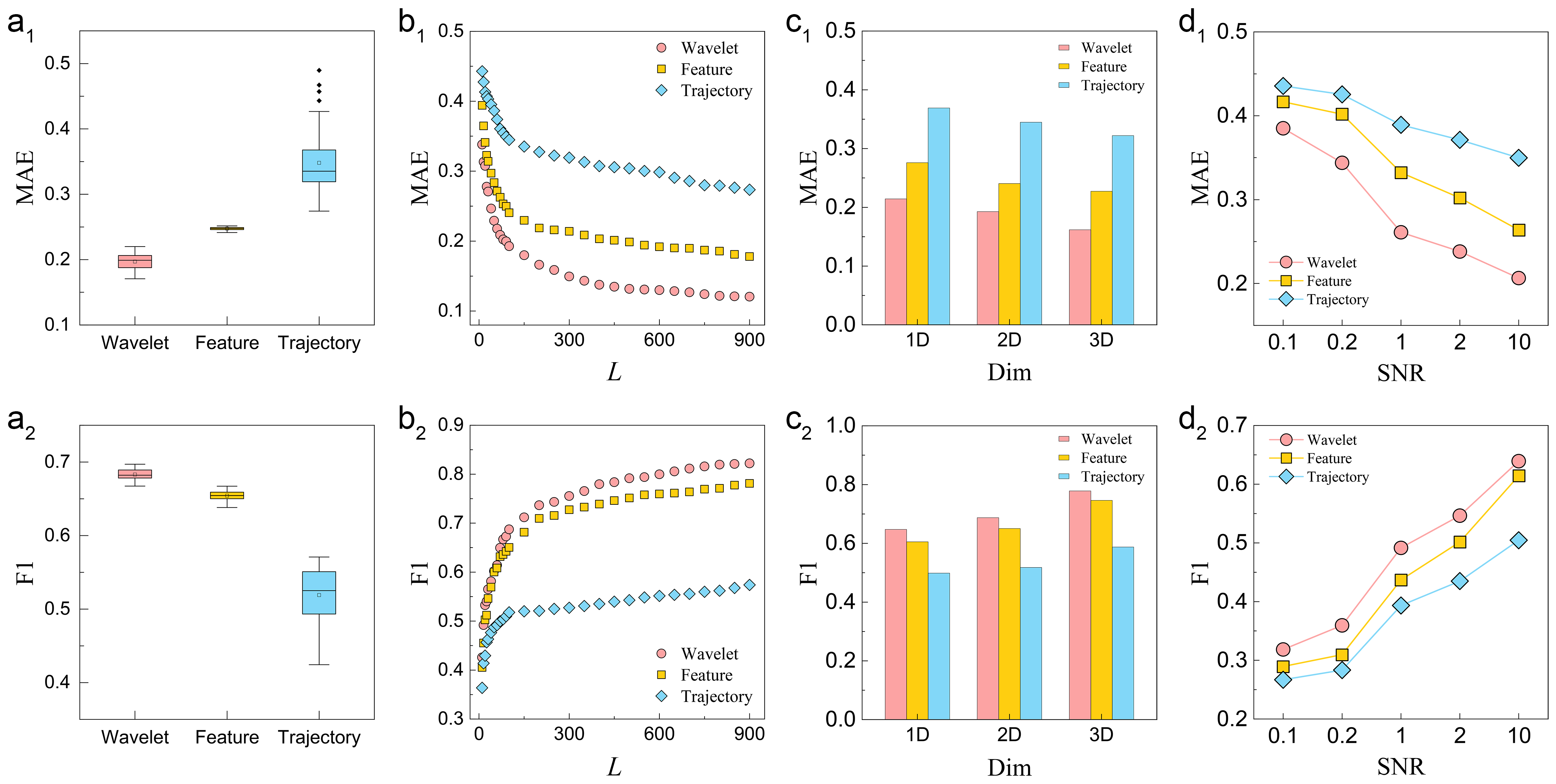}
\caption{Detailed performance analysis on simulated trajectories in the small-data regime ($N_{\rm train}=1000$). (a$_1$, a$_2$) Boxplots of test MAE and F1 scores over 50 independent training subsets. Wavelet- and feature-based models exhibit high stability (tight distributions), contrasting with the high variance of trajectory-based models. (b$_1$, b$_2$) Dependence on trajectory length $L$; the wavelet representation outperforms the baselines across all lengths. (c$_1$, c$_2$) MAE and F1 scores for 1D, 2D, and 3D trajectories, confirming consistent performance rankings across spatial dimensions. (d$_1$, d$_2$) Performance versus signal-to-noise ratio (SNR); the wavelet representation demonstrates higher noise resilience compared to feature- and trajectory-based alternatives.}
\label{fig:fig5}
\end{figure*}

The experimental SPT dataset in this work originates from the study by N. Granik et al. \cite{Granik2019}, which contains fluorescence trajectories of micron-sized beads embedded in reconstituted F-actin networks with controlled mesh sizes. In their experiments, fluorescent polystyrene beads with diameter $d_{\rm PS}\approx550~{\rm nm}$ are dispersed in entangled F-actin gels prepared at different actin monomer concentrations, yielding networks with characteristic mesh sizes $\xi \approx 225, 250, 300, 550, 625,$ and $750~\mathrm{nm}$. The mesh size for each condition is estimated from established microrheology calibrations. By tuning the mesh size, the authors modulate the crowding and viscoelastic properties of the medium, thereby altering the diffusive behavior of beads. 2D trajectories are then acquired by video fluorescence microscopy. Illustrations of different mesh sizes and their corresponding representative trajectories in this dataset are shown in Fig. \ref{fig:fig3}(a).

The SPT trajectories in this dataset exhibit substantial heterogeneity in length across the six mesh sizes. As reported in Appendix \ref{SPTdata}, for the three smaller meshes ($225$, $250$, and $300~\mathrm{nm}$), most trajectories contain more than $10^3$ time steps, with a broad tail extending up to several thousand steps. For the larger meshes ($550$, $625$, and $750~\mathrm{nm}$), the trajectories are typically shorter, with lengths concentrated in the range of a few hundred time steps. Moreover, to characterize the dynamical behavior under each condition, we estimate the diffusion exponent $\alpha$ for every trajectory by fitting the slope of its time-averaged mean-squared displacement (TA-MSD) in log-log coordinates. The resulting distributions of $\alpha$ are shown in Fig.~\ref{fig:fig3}(b). For all six mesh sizes, the $\alpha$ values are broadly distributed around a well-defined peak. As the mesh size increases from $225$ to $750~\mathrm{nm}$, the peak of the distribution shifts toward larger $\alpha$, with mean exponents (highlighted by red dashed lines) increasing from clearly subdiffusive values below unity at small meshes to values close to $\alpha \approx 1$ at large meshes.

To adapt these experimental trajectories to supervised learning, we construct fixed-length segments and define two learning tasks. For each mesh size $\xi$, the pool of available long trajectories is first randomly split into disjoint training, validation, and test sets. After this trajectory-level separation, long tracks are partitioned into non-overlapping 2D segments of a target length $L \in \{50,100,150,200,250\}$. All segments are then standardized (zero mean, unit variance per segment) and randomly shuffled within each subset. Each segment inherits the mesh-size label of its parent trajectory as well as the diffusion exponent $\alpha$ estimated from its TA-MSD. Based on these labels, two supervised tasks are considered, in direct analogy with the simulated case in Sec.~\ref{task1}:
\begin{enumerate}
\item The first is {\em diffusion-exponent regression}, where the target is the TA-MSD-based exponent $\alpha$ of each segment. Model performance is quantified by the MAE, defined as in Eq. \eqref{MAE}.
\item The second is {\em mesh-size classification}, in which each segment is assigned to one of the mesh-size conditions $\xi \in \{225,250,300,550,625,750\}\,\mathrm{nm}$. Performance is evaluated using the micro-averaged F1 score, following the definition in Eq. \eqref{F1}.
\end{enumerate}

\section{Results on simulated trajectories}\label{simu-results}

In this section, we systematically compare the proposed wavelet representation with feature-based and trajectory-based representations on the simulated benchmark described in Sec. \ref{task1}. The detailed choice of features used in the feature-based representation can be found in Table 3 of Ref. \cite{cai2025machine}, whereas the trajectory-based representation is obtained by applying per-trajectory standardization to the raw trajectories. For each representation family, we couple the input encoding with strong supervised learners (feature vectors with decision-tree or linear models, raw trajectories with deep sequence networks, and wavelet representations with modern vision architectures). The specific ML models used for each representation, together with the abbreviations adopted in this work, are summarized in Table \ref{tab:mlmodels}.

First, we evaluate these learners on diffusion-exponent regression (MAE) and diffusion-model classification (micro-F1) for 2D simulated trajectories of fixed length $L=100$, across a range of training set sizes $N_{\rm train}$. Here, Figs. \ref{fig:fig4}(a) and \ref{fig:fig4}(b) present a comprehensive performance comparison of ML models listed in Table~\ref{tab:mlmodels} across a wide spectrum of training set sizes, ranging from $10^3$ to $10^6$ trajectories. Each heatmap reports test set scores (on $2\times10^5$ trajectories) as a function of $N_{\rm train}$. Across this entire range and for both tasks, models built on the wavelet representation consistently achieve lower MAE and higher F1 than those using feature-based or trajectory-based representations. This advantage is particularly pronounced in the small-data regime, but it remains visible even when the training set contains as many as $10^6$ simulated trajectories.

In addition, the comparison between feature-based and trajectory-based representations highlights a distinct trade-off regarding data scalability. With limited training data, feature-based models generally outperform trajectory-based deep sequence models, reflecting the benefit of explicit physics-informed descriptors in the low-sample regime. However, their performance quickly saturates as the dataset size increases. As the training set grows ($N_{\rm train} \geq 10^5$), the trajectory-based models gradually narrow the gap and eventually outperform the feature-based ones, highlighting their higher asymptotic capacity when sufficient data are available. Nevertheless, they remain clearly inferior to the wavelet-based approach.

Moreover, to further illustrate data efficiency, Figs. \ref{fig:fig4}(c) and \ref{fig:fig4}(d) depict the performance distribution of individual models at a fixed training size of $N_{\rm train}=1000$. A clear hierarchy emerges that underscores the data-efficient nature of the wavelet representation: wavelet-based models cluster in the high-performance region (low MAE, high F1), followed by feature-based models, while trajectory-based models struggle to generalize from such limited data.

Based on these comparative results, we establish a standardized setup for the subsequent experiments, which focus specifically on the challenging small-data regime ($N_{\rm train}=1000$). We select the top-performing model from each representation family to serve as a representative baseline for further comparison:
\begin{enumerate}
\item EfficientNet (EffNet) is selected to represent the {\em wavelet} approach.
\item LightGBM (LGB) is chosen as the representative for {\em feature}-based learning.
\item Transformer (Trm) is selected to represent {\em trajectory}-based sequence modeling.
\end{enumerate}
Unless otherwise specified, these three models will serve as the benchmarks for their respective representation families in the remainder of this work.

Having established the baseline performance, we now analyze the three representation families in more detail in the small-data regime ($N_{\rm train}=1000$). We examine four aspects: stability against sample variation, dependence on trajectory length, generalization across spatial dimensionality, and robustness to localization noise, with the results summarized in Fig.~\ref{fig:fig5}. We begin by assessing the stability of each representation under random resampling of the training set. To this end, we construct 50 independent training subsets of size $N_{\rm train}=1000$ and evaluate the models on a fixed test set. The resulting boxplots in Figs. \ref{fig:fig5}(a$_1$) and \ref{fig:fig5}(a$_2$) reveal a clear contrast: while the wavelet- and feature-based models exhibit tight, compact distributions, the trajectory-based model suffers from significant variance and numerous outliers. This result underscores that deep sequence models become highly sensitive to sample selection when constrained by limited data, whereas the wavelet representation provides a robust embedding that is largely invariant to data resampling.

We then characterize the dependence on trajectory length $L$ and spatial dimensionality. As shown in Figs. \ref{fig:fig5}(b$_1$) and \ref{fig:fig5}(b$_2$), although performance naturally improves with increasing $L$ across all methods due to the accumulation of temporal information, the wavelet representation maintains a decisive lead throughout the entire range. This advantage extends to the spatial domain as well; Figs. \ref{fig:fig5}(c$_1$) and \ref{fig:fig5}(c$_2$) confirm that the performance hierarchy is preserved across 1D, 2D, and 3D systems, indicating that the efficacy of the wavelet approach is intrinsic to the representation rather than an artifact of a specific dimensional setting. Finally, we investigate the robustness of the representation to localization noise. We add zero-mean Gaussian noise with standard deviation $\sigma_n$ to the simulated trajectories, corresponding to signal-to-noise ratios $\mathrm{SNR}=1/\sigma_n$, and evaluate performance across a range of noise levels. As shown in Figs.~\ref{fig:fig5}(d$_1$) and \ref{fig:fig5}(d$_2$), while performance inevitably degrades for all models as the SNR decreases, the wavelet-based model remains markedly more resilient, maintaining the lowest MAE and highest F1 even in the noisiest conditions, whereas trajectory-based models deteriorate sharply. This suggests that the wavelet transform effectively filters high-frequency fluctuations while preserving the essential multiscale features required for diffusion analysis.

This comprehensive superiority is further corroborated by a detailed comparison across individual diffusion models (ATTM, CTRW, FBM, LW, and SBM), presented in Appendix \ref{simmodels}. Irrespective of the underlying physical mechanism, the wavelet representation consistently achieves the lowest regression errors and highest classification scores, confirming that its advantage is universal across the different types of anomalous diffusion considered. Collectively, these findings on simulated trajectories demonstrate that the wavelet representation is more data-efficient and robust than existing alternatives, providing a strong rationale for its application to experimental trajectories where data scarcity and noise are unavoidable.

\section{Results on experimental SPT trajectories}\label{exp-results}

Building on the analysis of simulated benchmarks, we now address the critical challenge of learning directly from real experimental trajectories. In this section, we compare the three representation families (wavelet-, feature-, and trajectory-based representations) on SPT recordings of fluorescent beads diffusing in F-actin networks, employing EffNet, LGB, and Trm as their representative learners. Distinct from the simulation-based paradigm that leverages large synthetic datasets, our approach here relies exclusively on experimental trajectories for both training and evaluation. To this end, we deliberately restrict the number of labeled tracks to mimic the data-scarce and noise-dominated conditions characteristic of real SPT experiments.

To rigorously evaluate the representations in the data-scarce regime, we curate small experimental subsets that closely mirror the simulated conditions in Sec. \ref{simu-results}. For each of the six mesh-size conditions ($\xi \in \{$225, 250, 300, 550, 625, 750$\} ~{\rm nm}$), we randomly sample 200 trajectories to form a balanced training set, yielding a total of $N_{\rm train}=1200$ experimental tracks. This scale aligns with the small-data regime ($N_{\rm train}=1000$) investigated in the simulated benchmarks. For performance evaluation, we construct a separate, larger test set by sampling 800 independent trajectories per condition (totaling $N_{\rm test}=4800$). To assess the robustness of each representation across different trajectory durations, we repeat this sampling procedure for five distinct segment lengths $L \in \{$50,100,150,200,250$\}$. The ensuing analysis is organized into two parts. In Subsec. \ref{exp-direct}, we benchmark the three representation families directly on these limited experimental datasets. Subsequently, in Subsec.~\ref{exp-sim}, we challenge the prevailing simulation-based paradigm by contrasting the wavelet-based model trained only on the small experimental set against SOTA deep networks trained on massive simulated trajectories.

\subsection{Benchmarking representations via direct learning from limited experimental trajectories}\label{exp-direct}

\begin{figure}
\centering
\includegraphics[width=8.5cm]{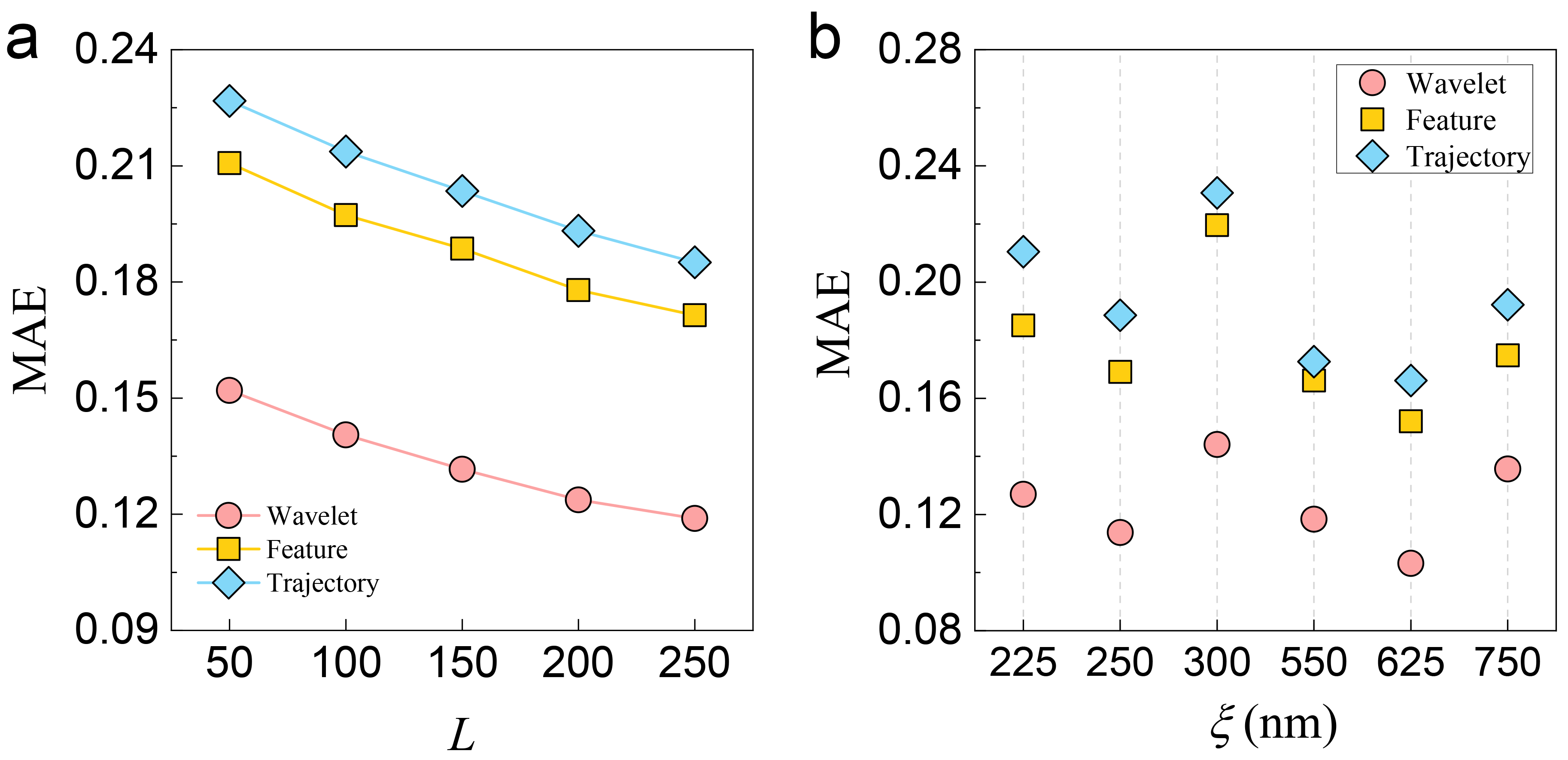}
\caption{Results of diffusion-exponent regression on experimental SPT trajectories in F-actin networks with $N_{\rm train}=1200$. (a) Test MAE as a function of segment length $L$ for wavelet-, feature-, and trajectory-based representations; the wavelet representation consistently yields the lowest error across all lengths. (b) Test MAE resolved by mesh size $\xi$ at $L=200$, showing that the wavelet representation maintains the smallest error for all network conditions.}
\label{fig:fig6}
\end{figure}

\begin{figure*}
\centering
\includegraphics[width=17.8cm]{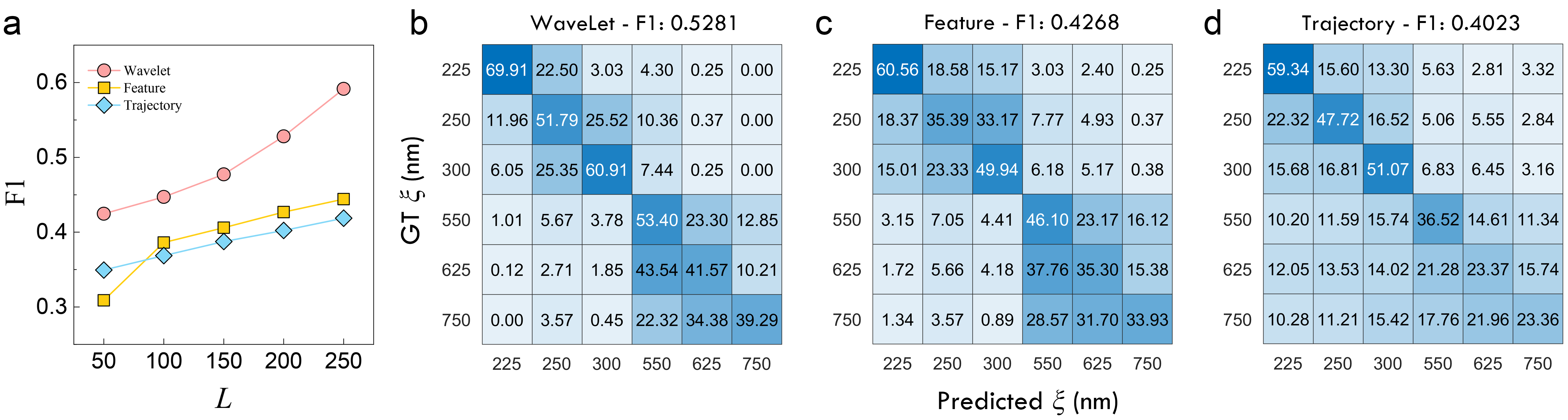}
\caption{Mesh-size classification on experimental SPT trajectories in F-actin networks with $N_{\rm train}=1200$. (a) Micro-averaged F1 score as a function of segment length $L$ for wavelet-, feature-, and trajectory-based representations; the wavelet representation achieves the highest F1 for all $L$. (b)-(d) Confusion matrices for $L=200$ for the wavelet, feature-based, and trajectory-based representations, respectively [rows: ground-truth (GT) mesh size $\xi\in\{225,250,300,550,625,750\}~\mathrm{nm}$; columns: predicted $\xi$; entries in percent].}
\label{fig:fig7}
\end{figure*}

In this subsection, we first benchmark the three representation families on the diffusion-exponent regression and mesh-size classification tasks defined in Sec. \ref{task2}, using the small experimental training and test sets described above ($N_{\rm train}=1200$, $N_{\rm test}=4800$).

For the diffusion-exponent regression, Fig. \ref{fig:fig6}(a) reports the MAE as a function of segment length $L$. While prediction error naturally decreases with $L$ for all methods, the wavelet representation establishes a distinct advantage, achieving the lowest MAE across the entire range. Notably, the feature-based representation, although strong on simulated benchmarks, exhibits a pronounced gap in this experimental setting. This suggests that with only 1200 training samples, handcrafted statistical descriptors suffer from fluctuations inherent to experimental noise, whereas the wavelet transform extracts more robust signatures. The trajectory-based model yields the largest errors, reinforcing the difficulty of training deep sequence models on raw coordinates without large datasets. The robustness of the wavelet representation is further highlighted in Fig. \ref{fig:fig6}(b), which decomposes the MAE by mesh size $\xi$ at $L=200$. Regression difficulty varies across mesh sizes, and all models show elevated errors at specific conditions (e.g., $\xi=300$ and $750~\mathrm{nm}$), yet the wavelet representation consistently attains the lowest MAE. This indicates that the wavelet representation provides a robust basis for diffusion analysis that is largely insensitive to the specific viscoelastic properties of the environment.

In the mesh-size classification task, discerning the exact environmental condition proves challenging in the small-data regime due to the small variations between mesh sizes (as small as 25 nm) and the presence of experimental noise. Consequently, absolute F1 scores remain moderate for all representations (Fig. \ref{fig:fig7}). Nevertheless, the wavelet representation establishes a clear lead over the alternatives. Fig. \ref{fig:fig7}(a) shows that while the wavelet representation's F1 score scales naturally with $L$, reaching $\approx 0.6$ at $L=250$, the feature-based and trajectory-based representations plateau at lower levels (F1 $\approx 0.40-0.45$). The confusion matrices for $L=200$, presented in Figs. \ref{fig:fig7}(b)-(d), offer physical insight into this disparity.

The experimental system involves beads of diameter $d_{\rm PS} \approx 550 {\rm nm}$ diffusing in meshes of varying sizes. For the smaller meshes ($\xi\in\{225,250,300\}~\mathrm{nm}$), where $\xi<d_{\rm PS}$, the beads are strongly confined. In this regime, all three representations capture the signatures of confinement relatively well, as evidenced by the recognizable diagonal structures in the top-left $3\times 3$ blocks of the confusion matrices. However, the wavelet representation [Fig. \ref{fig:fig7}(b)] yields the sharpest separation with the least leakage between adjacent classes, indicating that it successfully resolves the subtle variations in confinement degree that feature- and trajectory-based statistics miss. For the larger meshes ($\xi\in\{550,625,750\}~\mathrm{nm}$), the mesh size becomes comparable to or larger than the bead diameter. In this regime, the bead motion is less strongly constrained and relatively closer to free diffusion, making the trajectories more statistically similar and difficult to differentiate. Reflecting this physical ambiguity, both wavelet and feature-based models [Figs. \ref{fig:fig7}(b) and (c)] exhibit ``blocky'' confusion regions among these large-mesh classes, indicating that distinguishing between these subtle variations in mesh size is challenging for any method. Crucially, however, the trajectory-based model [Fig. \ref{fig:fig7}(d)] suffers from a more severe failure mode: it displays significant off-diagonal dispersion that extends beyond the large-mesh block, frequently misclassifying large-mesh trajectories (less confined) as belonging to small-mesh conditions (confined). This contrast highlights that while the large-mesh regime is difficult for all, the wavelet representation remains the most robust, preserving the distinction between confined and less-confined states better than the alternatives.

\subsection{Wavelet representations versus the simulation-based paradigm}\label{exp-sim}

As highlighted in Sec. \ref{introduction}, the general strategy for the machine-learning analysis of anomalous diffusion relies on the simulation-based paradigm. In this paradigm, ML models are trained on large synthetic datasets and then transferred to experiments. Despite exploiting effectively unlimited labeled trajectories, their performance is ultimately constrained by the simulation-to-reality gap, because idealized diffusion models cannot fully reproduce the complex noise patterns, viscoelastic memory, and transient heterogeneities present in real recordings. Consequently, simply increasing the volume of synthetic training data yields diminishing returns once performance is limited by this distributional mismatch rather than sample size. In this subsection, we challenge this paradigm by conducting a direct head-to-head comparison. We contrast the performance of the WADNet, a SOTA deep model from the AnDi Challenge, trained on up to $10^6$ simulated trajectories, against our wavelet-representation-based model (EffNet), trained directly on a small set of 1200 experimental segments.

\begin{figure*}
\centering
\includegraphics[width=17.0cm]{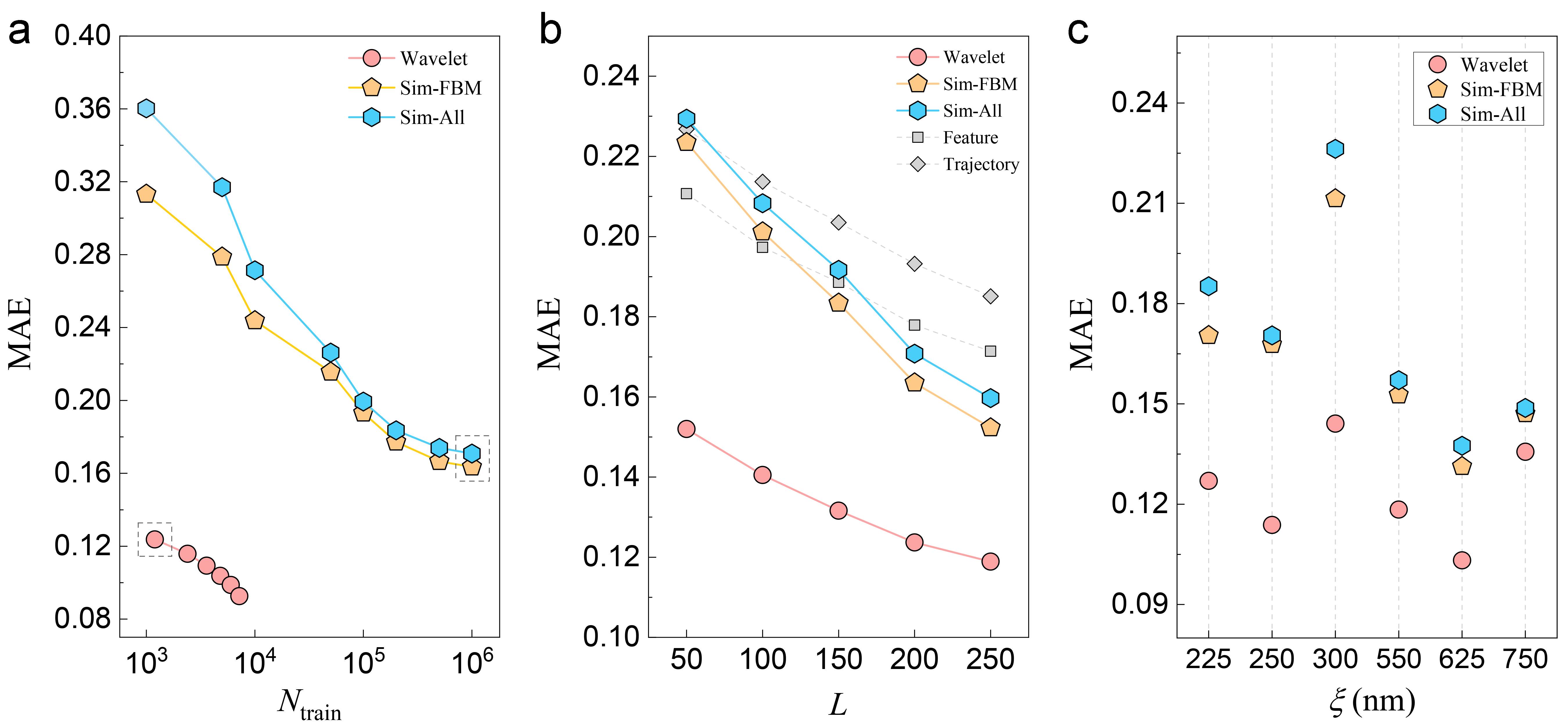}
\caption{Wavelet representations versus the simulation-based paradigm on the experimental SPT test set ($N_{\rm test}=4800$). (a) MAE for the diffusion-exponent regression at fixed segment length $L=200$ as a function of training set size $N_{\rm train}$. Dashed boxes highlight the training regimes used for the comparisons in (b) and (c). (b) MAE as a function of segment length $L$ for the wavelet representation, Sim-FBM, Sim-All, feature-, and trajectory-based models. Across all lengths, the wavelet representation attains the lowest MAE. (c) MAE resolved by mesh size $\xi$ for $L=200$, where the wavelet model achieves the smallest error for all network conditions.}
\label{fig:fig8}
\end{figure*}

To push this simulation-based baseline to its performance limit, we construct two optimized synthetic training sets for the WADNet, each expanded to $10^6$ trajectories to represent the saturation level of the paradigm:
\begin{enumerate}
\item Sim-FBM: Consisting exclusively of FBM trajectories. Since the original experimental study \cite{Granik2019} identified FBM as the primary diffusion mechanism of the beads, this configuration provides the simulation model with strong physical priors, effectively creating an ``idealized'' training scenario informed by prior literature.
\item Sim-All: A balanced mixture of the five standard anomalous diffusion models (ATTM, CTRW, FBM, LW, SBM) to promote generalization and reduce potential model bias.
\end{enumerate}
Crucially, to reduce the domain shift between simulation and reality, we add zero-mean Gaussian noise into these simulated tracks. The noise intensity $\sigma_n$ is selected by grid search to best match the localization error of the experimental setup (see Appendix \ref{addnoise}). These massive, noise-matched simulated trajectories are designed to approximate an upper bound on the performance achievable within the simulation-based paradigm.

Before proceeding to the formal comparison, we first assess whether performance has saturated with respect to training data by examining how it scales for the two competing approaches at $L=200$, as illustrated in Fig. \ref{fig:fig8}(a). For the simulation-based models, increasing the synthetic training set size from $10^4$ to $10^6$ trajectories yields only marginal improvements on the experimental test set, leading to a clear performance plateau. This observation indicates that the performance of WADNet is limited by a persistent mismatch between simulated and real diffusion dynamics, rather than by a lack of training samples. By contrast, the wavelet representation shows no sign of saturation. As the experimental training set size increases from $N_{\mathrm{train}}=1200$ to $N_{\mathrm{train}}=7200$, the prediction error consistently decreases, indicating that the model has not yet reached its performance ceiling. Consequently, the comparison presented below pits the approximately saturated performance of the simulation-based paradigm against the early-stage, data-constrained performance of the wavelet representation, with the specific comparison regimes highlighted by dashed boxes in Fig. \ref{fig:fig8}(a).

With the saturation behavior established, we now turn to the direct performance comparison. Fig. \ref{fig:fig8}(b) presents the diffusion-exponent regression results for the WADNet trained on the large $N_{\mathrm{train}}=10^6$ simulated datasets versus the experimental models trained on the restricted $N_{\mathrm{train}}=1200$ subset. The wavelet representation consistently attains the lowest MAE, significantly outperforming both the Sim-FBM and Sim-All baselines across all trajectory lengths. Despite the three orders of magnitude disparity in training data volume, the wavelet representation extracts more relevant physical information from a few thousand real segments than the deep learning baselines learn from millions of imperfectly matched simulated trajectories. This superiority is robustly preserved across diverse physical environments, as shown in the mesh-size decomposition in Fig. \ref{fig:fig8}(c) (for $L=200$). Regardless of the specific viscoelastic condition, the wavelet representation maintains the lowest error, whereas Sim-FBM and Sim-All consistently yield larger errors by a visible margin. It is worth noting that in the experimental context, the ground truth value of $\alpha$ is operationally defined by the TA-MSD fit rather than a theoretical generation parameter. Therefore, the superior performance of the wavelet representation reflects, at least in part, its ability to reproduce this established experimental standard, effectively bypassing the simulation-to-reality gap that limits simulation-trained models.

However, attributing this superiority solely to the elimination of the simulation-to-reality gap would be an oversimplification. The performance of feature- and trajectory-based representations trained on the same 1200 experimental segments provides a critical test of this hypothesis. Theoretically, direct experimental learning benefits from inherent label consistency: the training targets (TA-MSD estimates) share the same distribution, noise characteristics, and estimator biases as the test set, whereas simulation-based training relies on theoretical ground-truth labels that may not perfectly map to empirical estimators. Yet, despite this intrinsic advantage of in-distribution training, Fig. \ref{fig:fig8}(b) shows that the feature- and trajectory-based models yield error rates on par with the simulation-based WADNet, failing to replicate the substantial gain achieved by the wavelet model. This finding is pivotal: it strongly suggests that the performance leap cannot be attributed solely to in-distribution training on experimental data. Instead, it confirms that the breakthrough relies on the specific synergy between the on-distribution nature of direct experimental learning and the high data efficiency of the wavelet representation. Collectively, these results demonstrate that the wavelet representation constitutes a viable alternative to the simulation-based paradigm, enabling high-precision analysis by learning directly from limited experimental recordings without relying on large simulated datasets.

\section{Interpretability: The Physical Origin of Data Efficiency}\label{interpretability}

\begin{figure*}
\centering
\includegraphics[width=17.5cm]{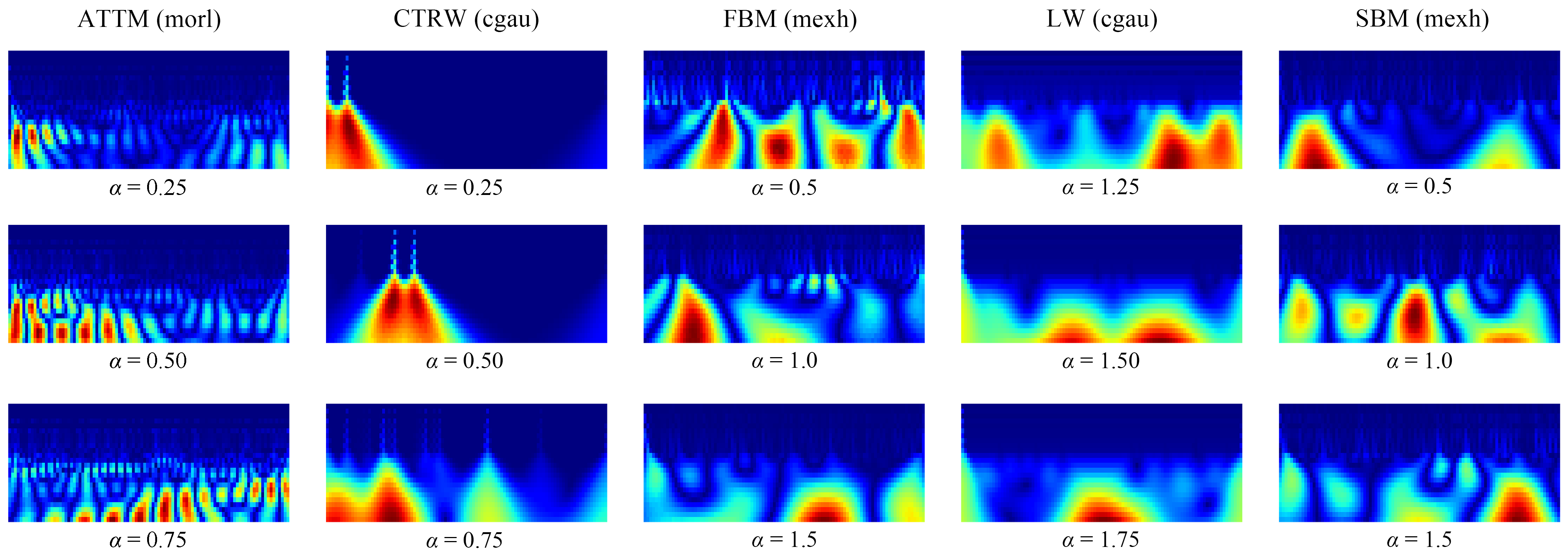}
\caption{Representative wavelet scalograms (trajectory length $L=100$) for five anomalous diffusion models (columns) under varying anomalous exponents $\alpha$ (rows). To visualize the most discriminative features, each model is paired with its optimal mother wavelet (noted in parentheses), selected based on the highest single-wavelet classification accuracy (see Appendix \ref{waveletselection}).}
\label{fig:fig9}
\end{figure*}

The superior data efficiency and generalization capability of the wavelet representation, as demonstrated in the preceding sections, warrant a physical explanation. Why does transforming a trajectory into a time-scale scalogram allow a machine learning model to learn more effectively from limited data than utilizing the raw trajectory or handcrafted features? Our central thesis is that the CWT explicitly disentangles the multiscale physical mechanisms driving anomalous diffusion, such as non-stationarity, temporal correlations, and transient heterogeneities. These mechanisms are often hidden in the time domain or averaged out by global statistical descriptors, whereas the wavelet scalogram renders them explicit as distinct geometric patterns in the time-scale plane.

To understand the physical content of the scalograms, we first interpret the continuous wavelet transform in the language of stochastic processes. For a trajectory $x(t)$, the coefficient $W_\psi(a,b)$ at scale $a$ and time $b$ is the inner product between $x(t)$ and a time-localized analyzing wavelet centered at $b$ with an effective support of order $a$. Its modulus $|W_\psi(a,b)|$ therefore measures the magnitude of fluctuations of $x(t)$ on time scales of order $a$ in a neighbourhood of $b$. In this sense, the scalogram $|W_\psi(a,b)|$ can be regarded as a \emph{local multiscale structure function}: at each point $(a,b)$ it probes the amplitude of fluctuations on a characteristic time scale $a$.

As a representative example, standard results from wavelet analysis for self-similar processes show that the moments of coefficients scale as $\langle |W_\psi(a,b)|^q\rangle_b \propto a^{\zeta(q)}$ \cite{muzy1991wavelets}. Specifically, the scaling of the modulus closely tracks the diffusion exponent $\alpha$ [e.g., $\zeta(2)=\alpha+1$ for FBM]. Thus, the vertical profile of the scalogram encodes information equivalent to the MSD, but resolved locally. Crucially, unlike the global MSD which averages over time, the time localization of the CWT preserves non-stationary features, such as diffusivity changes or trapping events. This capability creates distinct 2D ``scale fingerprints'' that reveal the underlying physical mechanisms, which we examine in detail below for both simulated diffusion models and experimental SPT trajectories.

\subsection{Scale fingerprints of anomalous diffusion models}\label{simspectra}

To illustrate how these scale fingerprints manifest in practice, we examine the wavelet spectra of the five prototypical models implemented in the andi-datasets library. As our machine learning pipeline utilizes a stack of six complementary wavelet families, a specific criterion is required to select the most appropriate channel for visual interpretation. Guided by the ablation study in Appendix \ref{waveletselection}, we adopt a best-in-class strategy for this purpose: for each diffusion model, we select the mother wavelet that attains the highest diagonal classification accuracy for that class in the single-wavelet setting. Consequently, the representative scalograms are generated using the morl for ATTM, cgau for CTRW and LW, and mexh for FBM and SBM, which are presented in Fig. \ref{fig:fig9} (with trajectory length $L=100$). Although the raw trajectories often appear qualitatively similar, their wavelet spectra display highly structured, model-specific fingerprints driven by distinct physical mechanisms underlying each type of anomalous diffusion:

\begin{enumerate}
    \item ATTM (From diffusivity switching to patchy patterns):
    In the ATTM model, the particle undergoes Brownian motion with a diffusion coefficient $D(t)$ that resets stochastically. Within a constant-$D$ interval, the process exhibits stationary increments; however, sharp transitions in $D$ cause simultaneous magnitude jumps across all scales. This mechanism manifests in the scalogram as a ``patchy'' texture (Fig. \ref{fig:fig9}, first column), where rectangular blocks of uniform intensity are separated by vertical edges. For classification, these edges serve as a distinct signature of medium heterogeneity. Regarding parameter estimation, the contrast between these blocks can be qualitatively associated with the heterogeneity of the diffusion process in our simulations; as $\alpha$ increases (from top to bottom), the joint statistics of diffusivities and residence times change so that the texture tends to become more uniform and less segmented. This gradual homogenization allows the vision models to regress $\alpha$ based on the decaying prominence of the patches.

    \item CTRW (From trapping events to vertical gaps):
    The subdiffusive CTRW is governed by heavy-tailed waiting times $\Psi(t) \sim t^{-\sigma}$ where diffusion exponent $\alpha = \sigma-1$. During a trapping event, the particle is effectively immobilized, which suppresses wavelet coefficients across all scales. In the scalogram (Fig. \ref{fig:fig9}, second column), these events appear as deep vertical ``gaps'' or cones of silence. This sparse pattern is the primary cue for diffusion-model classification. Crucially for exponent regression, the statistics of the width and density of these gaps depend on the anomalous exponent $\alpha$ through the waiting-time distribution. As shown in Fig. \ref{fig:fig9}, a lower $\alpha$ (top row, $\alpha=0.25$) corresponds to a heavier tail in the waiting time distribution, resulting in wider and more prominent gaps compared to the higher $\alpha$ case (bottom row, $\alpha=0.75$), where the texture becomes denser with more closely spaced activity cones.

    \item FBM (From self-similarity to spectral slopes):
    FBM is a Gaussian process characterized by long-range correlations defined by the Hurst exponent $H$. Unlike ATTM or CTRW, FBM scalograms (Fig. \ref{fig:fig9}, third column) are statistically homogeneous along the time axis but exhibit a strong dependence on the scale axis. The classification relies on this temporal uniformity. For regression, the model leverages the ``spectral slope'' along the vertical axis. A low $\alpha$ (subdiffusive, top row) implies negatively correlated increments, concentrating energy at fine scales (high frequencies) and creating a rough texture. Conversely, a high $\alpha$ (superdiffusive, bottom row) implies positive correlations, shifting energy to coarse scales (low frequencies) and resulting in a smoother texture with high-intensity regions at the bottom of the scalogram.

    \item LW (From ballistic flights to coarse-scale ridges):
    L\'evy walks consist of ballistic flights coupled with power-law distributed durations. A constant-velocity flight represents a linear trend, which manifests as an intense, horizontal ridge concentrated at coarse scales (Fig. \ref{fig:fig9}, fourth column). While the presence of these ridges distinguishes LW from the localized features of CTRW, their persistence encodes the diffusion exponent. As $\alpha$ increases towards the ballistic limit (comparing top row $\alpha=1.25$ to bottom row $\alpha=1.75$), the flight durations become longer. Consequently, the coarse-scale ridges in the scalogram become more continuous and elongated in time, providing a robust visual feature for estimating $\alpha$.

    \item SBM (From aging to temporal gradients):
    For SBM, the diffusivity scales deterministically as $D(t) \sim t^{\alpha-1}$. This aging process breaks time-translation invariance, creating a global gradient in scalogram intensity along the time axis (Fig. \ref{fig:fig9}, fifth column), which serves as the unique identifier for SBM. The direction and steepness of this gradient allow for precise regression of $\alpha$. For subdiffusive cases ($\alpha < 1$, top row), the diffusivity decays over time, causing the scalogram to fade from bright to dark. In contrast, for superdiffusive cases ($\alpha > 1$, bottom row), the diffusivity increases, resulting in a gradient that brightens from early to late times. This inversion of the temporal gradient provides a clear, monotonic cue for the vision models to determine the anomalous exponent.
\end{enumerate}

Collectively, these analyses make precise what we mean by ``scale fingerprints'': each physical mechanism gives rise to a characteristic arrangement of intensity in the time-scale plane, such as patchy domains from diffusivity switching, sparse gaps from heavy-tailed trapping, vertically homogeneous spectra with scale-dependent slopes from long-range correlations, coarse-scale ridges from ballistic flights, and temporal gradients from aging. In contrast to raw trajectories, where these mechanisms are often obscured by stochastic fluctuations, the wavelet representation disentangles them into a small set of robust multiscale patterns that are naturally amenable to pattern recognition by vision architectures. These fingerprints simultaneously encode model identity through their qualitative structure and the anomalous exponent through systematic deformations along the columns of Fig. \ref{fig:fig9}. Therefore, the wavelet representation reduces the effective complexity of the learning problem and enables vision models to achieve high accuracy in both diffusion-model classification and diffusion-exponent regression even in the small-data regime. In the next subsection, we show that analogous scale fingerprints also emerge in the experimental SPT trajectories in F-actin networks and that they carry clear signatures of confinement and mesh-size dependence.

\subsection{Experimental validation: Fingerprints of confinement in F-actin networks}\label{expspectra}

To validate whether the concept of scale fingerprints extends beyond theoretical prototypes to complex experimental systems, we analyze the SPT trajectories of fluorescent beads diffusing in F-actin networks. In this system, tracer diffusion is modulated by the mesh size $\xi$ of the network, which imposes steric constraints and yields a tunable degree of confinement. As in the previous subsection, we visualize single-channel wavelet spectra in order to highlight qualitative patterns. Given that previous studies \cite{Granik2019} have characterized these SPT trajectories as FBM, we employ the mexh to generate the experimental scalograms, as this kernel demonstrates superior efficacy for FBM in our simulated benchmarks (see Appendix \ref{waveletselection}).

Fig. \ref{fig:fig10} presents the wavelet spectra of experimental trajectory segments ($L=100$) with varying mesh sizes ranging from $\xi=225$ nm to $\xi=750$ nm. A visual inspection reveals that the mesh size acts as a control parameter that systematically alters the texture of the scalograms, mirroring the ``rough-to-smooth'' transition described in the theoretical FBM analysis (refer to Fig. \ref{fig:fig9}, third column). For dense networks with small mesh sizes $\xi \in \{225, 250, 300\}$ nm where $\xi < d_{\rm PS}$, the scalograms exhibit a ``rough'' and fragmented texture, particularly at fine scales. Physically, this texture reflects the strong steric confinement imposed by the dense network. Although the underlying mechanism is caging, the resulting rapid directional reversals phenomenologically resemble the antipersistent correlations of subdiffusive FBM. In the wavelet domain, these rapid reversals manifest as high-frequency interruptions that preclude the formation of continuous structures, resulting in a rough, vertically striated pattern. Note that, similar to the simulated FBM, these scalograms are approximately homogeneous along the time axis and do not exhibit pronounced long-time trends, distinguishing them from the transient patchiness of ATTM or the temporal gradients of SBM.

\begin{figure}
\centering
\includegraphics[width=8.6cm]{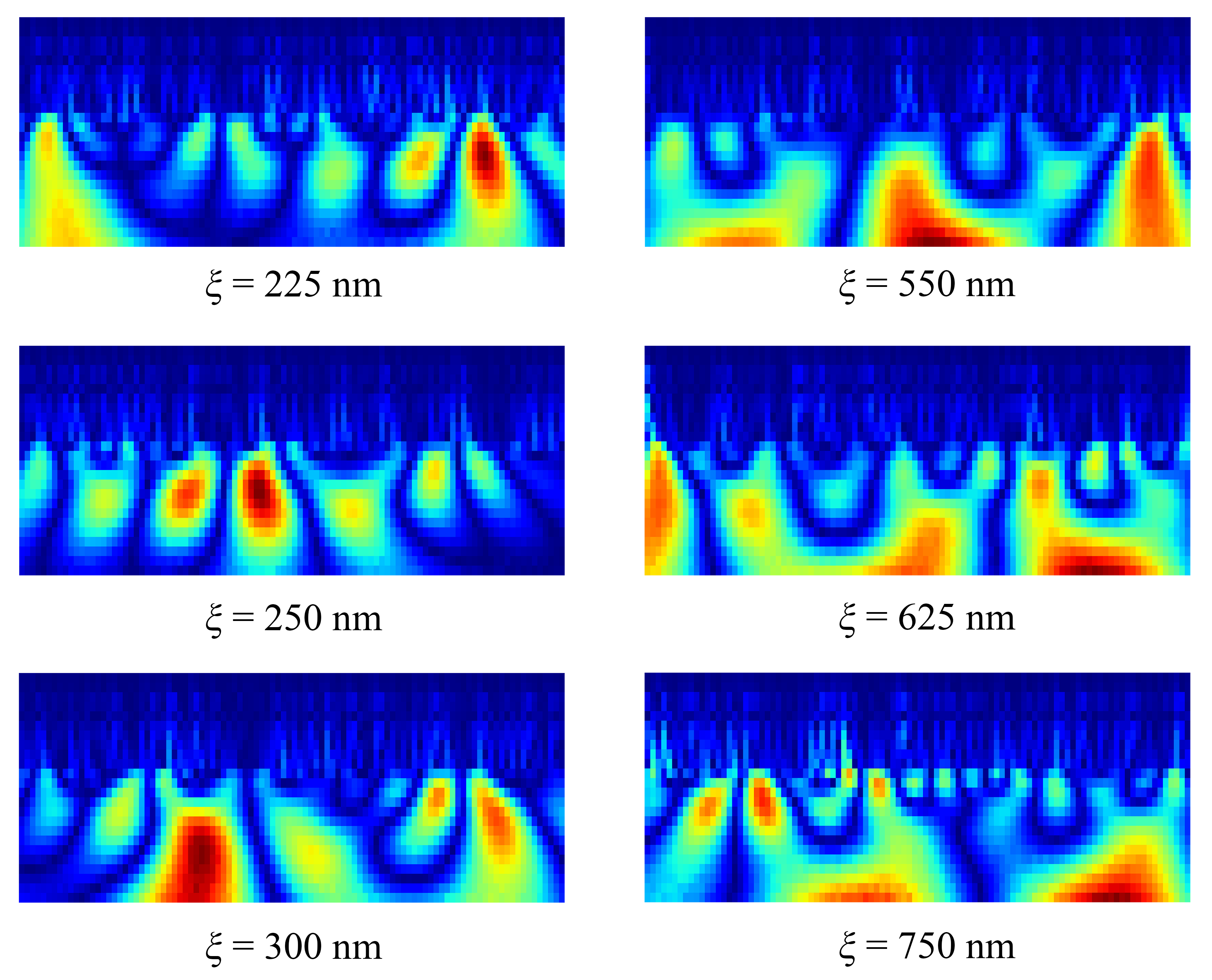}
\caption{Evolution of wavelet scalograms for experimental SPT trajectory segments ($L=100$) in F-actin networks with increasing mesh sizes $\xi$.}
\label{fig:fig10}
\end{figure}

As the mesh size increases to $\xi\in\{550,625,750\}$ nm where $\xi \geq d_{\rm PS}$, a morphological transition occurs. The distinct vertical striations gradually merge into smoother, continuous horizontal ridges that span extended temporal durations. At $\xi=750$ nm, the texture at the coarse scales (bottom of the scalogram) becomes significantly brighter and more coherent. This evolution is consistent with the relaxation of confinement constraints. As the pores enlarge, the bead motion approaches free diffusion, effectively reducing the apparent antipersistence observed at finer scales. This shift effectively redistributes spectral energy from fine to coarse scales, visually reproducing the signature of higher-$\alpha$ regimes.

These findings elucidate the physical origin of the wavelet representation's data efficiency. By converting subtle temporal correlations into salient, high-contrast visual features, the wavelet representation effectively lowers the learning barrier. The model does not need to implicitly derive the complex statistical laws of viscoelasticity from massive datasets of raw coordinates; instead, it can directly leverage these explicit textural motifs that robustly encode the system's state. This direct mapping from physical constraints (mesh size) to textural features (roughness vs. smoothness) significantly reduces the effective complexity of the learning problem, enabling the vision backbone to generalize accurately even when trained on limited examples.

\section{Conclusion}\label{conclusion}

In this work, we have introduced a wavelet-based representation of anomalous diffusion trajectories and demonstrated that, when combined with standard supervised models, it enables data-efficient learning directly from experimental recordings. By mapping trajectories to multi-channel wavelet modulus scalograms computed from six complementary continuous wavelet families, we obtain an image-like, multiscale encoding in which the temporal structure of random motion is rendered explicit in the time-scale plane. On simulated trajectories from the andi-datasets benchmark, models utilizing the wavelet representation maintain superior accuracy in diffusion-exponent regression and model classification with as few as 1000 training trajectories, while remaining robust to localization noise and variations in trajectory length. Most notably, when applied to experimental SPT data of fluorescent beads in F-actin networks, our approach achieves a decisive advantage. A vision model trained on merely 1200 experimental segments using the wavelet representation yields significantly lower regression errors than SOTA deep learning models trained on $10^6$ simulated trajectories. These results challenge the prevailing reliance on massive synthetic datasets as the primary route to high-performance ML for anomalous diffusion, suggesting instead that a physics-informed representation coupled with small-scale, in-distribution experimental data offers a more effective strategy.

This data efficiency of the wavelet representation originates from the ability of the continuous wavelet transform to explicitly disentangle the multiscale mechanisms governing random motion. As our interpretability analysis reveals, complex dynamic behaviors, such as diffusivity switching in ATTM, heavy-tailed trapping in CTRW, or antipersistent correlations in FBM, are rendered into distinct, stationary geometrical patterns or ``scale fingerprints'' in the time-scale plane. By converting temporal statistical properties into robust visual textures, the wavelet representation reduces the effective complexity of the learning problem, allowing standard vision architectures to identify physical states without implicitly reconstructing statistical laws from raw coordinates. The supporting codes in this work are accessible at our GitHub repository \cite{ourcodes}.

Looking forward, the wavelet representation opens several promising avenues for future research. One immediate avenue is the application of this framework to trajectory segmentation (or change-point detection) in heterogeneous environments. Given the wavelet transform's inherent sensitivity to local transient features (manifested as vertical edges or texture shifts in the scalograms), adapting object detection or segmentation networks to the wavelet domain could enable precise detection of state transitions even within extremely short trajectories, a regime where current statistical sliding-window methods \cite{Manzo2021,Bo2019,Argun2021,Dosset2016,Zhang2023,Wagner2017,Monnier2015,Matsuda2018,Gentili2021,Arts2019} often struggle. A second, perhaps more fundamental direction is to use this representation for the unsupervised discovery of novel diffusion behaviors. Current ML approaches are largely supervised, limiting analysis to a predefined set of theoretical models (e.g., the five standard models used in this work). However, experimental systems often exhibit complex dynamics that do not fit neatly into these categories. Since the wavelet representation provides a rich and distinct fingerprint for different dynamical universality classes, its combination with unsupervised clustering or manifold learning techniques could allow researchers to construct a purely data-driven taxonomy of random motion. Such an approach can enable the identification and characterization of previously unknown diffusion modes directly from experimental data, paving the way for a deeper understanding of transport phenomena in complex biological and soft matter systems.

\begin{acknowledgments}
We thank Wenjie Cai for helpful discussions. This work is supported by the National Natural Science Foundation of China (Grant No. 12104147) and the industry-university cooperation project with DataComp AI (Project No. H202591470640).
\end{acknowledgments}

\appendix

\section{Ablation study on the wavelet scale selection}\label{scaleselection}

\begin{figure}
\centering
\includegraphics[width=8.5cm]{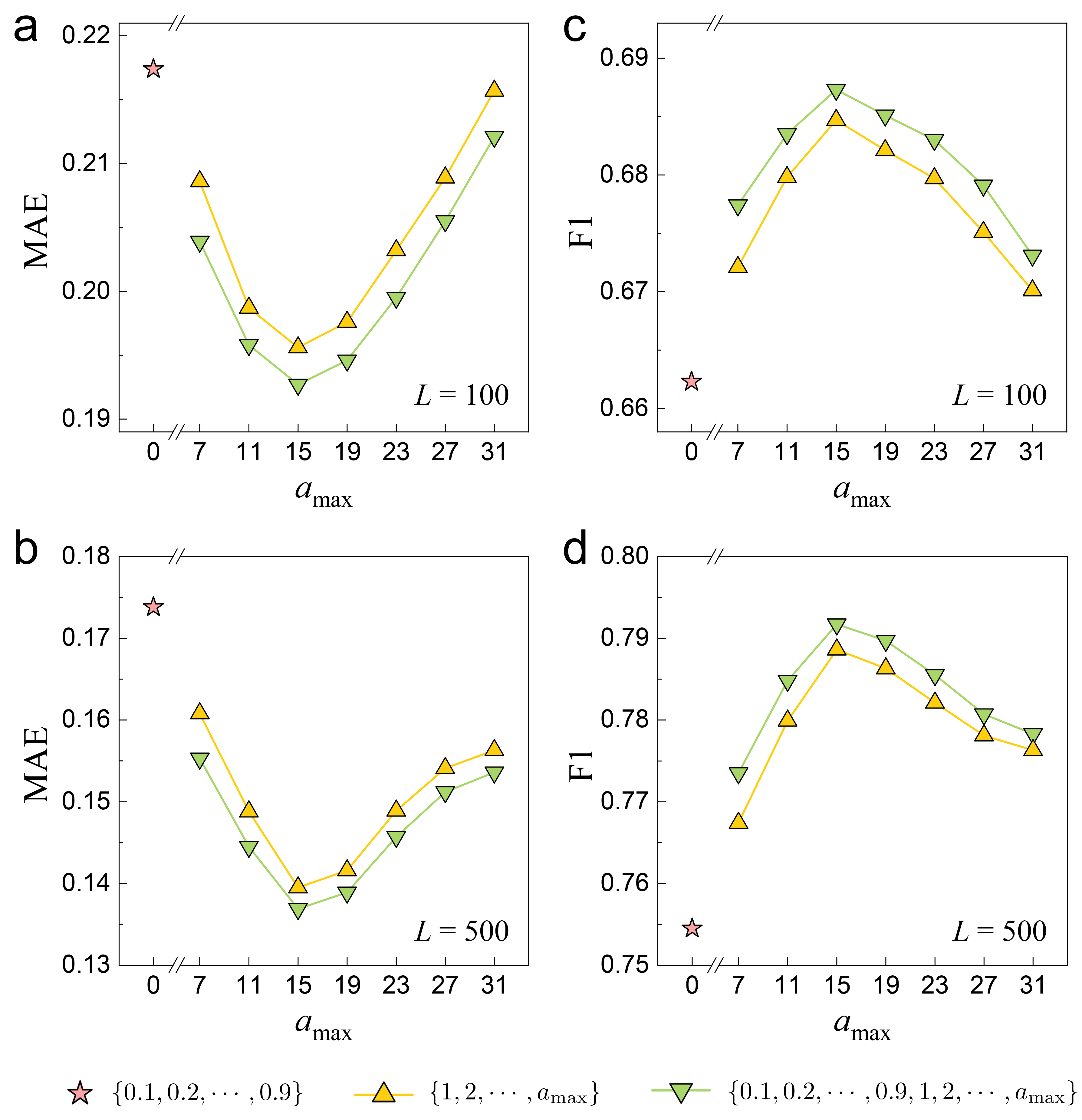}
\caption{Ablation study of the CWT scale selection on 2D simulated trajectories. (a) and (b) show the MAE of diffusion-exponent regression for trajectories of length $L=100$ and $L=500$, respectively, as a function of the maximum integer scale $a_{\max}$; (c) and (d) show the corresponding micro-F1 scores for diffusion-model classification. Both regression and classification performance are optimized around $a_{\max}=15$, and the combined scale consistently outperforms the integer-only one.}
\label{fig:figA1}
\end{figure}

\begin{figure*}
\centering
\includegraphics[width=14.5cm]{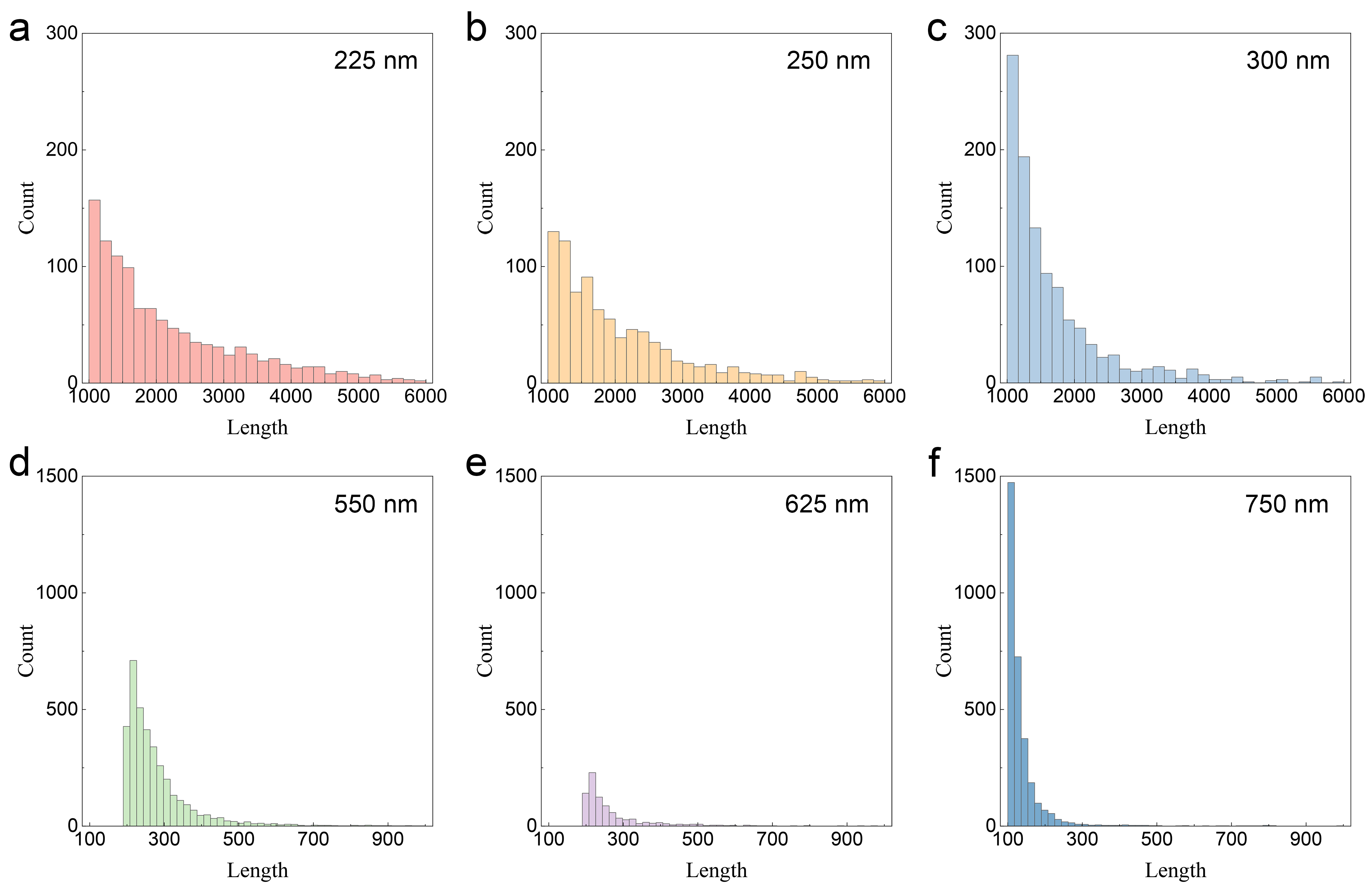}
\caption{Length distributions of experimental SPT trajectories in F-actin networks with mesh sizes $\xi = 225$ (a), $250$ (b), $300$ (c), $550$ (d), $625$ (e), and $750~\mathrm{nm}$ (f). For each condition, the histogram shows the number of recorded trajectories as a function of their length (number of time steps).}
\label{fig:figA2}
\end{figure*}

In the main text we fix the set of CWT scales to
$\{0.1,0.2,\ldots,0.9,1,2,\ldots,15\}$,
which combines a block of sub-unit scales with a finite range of integer scales. In this appendix we provide an ablation study that motivates this choice and, in particular, the upper cutoff at $a=15$ and the inclusion of the sub-unit scales $a<1$.

In this ablation study, 2D simulated trajectories of fixed length $L=100$ and $L=500$ generated with the andi-datasets library are considered. As in the main text, we evaluate the wavelet representation on two benchmark tasks: diffusion-exponent regression (MAE) and diffusion-model classification (micro-F1), using a training set of 1000 trajectories together with a test set of size $2\times10^5$. We vary the maximum integer scale $a_{\max}$ over the values
$a_{\max} \in \{7,11,15,19,23,27,31\}$, and compare three scale configurations:
\begin{enumerate}
\item sub-unit-only scale, $\{0.1,0.2,\ldots,0.9\}$;
\item integer-only scale, $\{1,2,\ldots,a_{\max}\}$;
\item combined scale, $\{0.1,0.2,\ldots,0.9,1,2,\ldots,a_{\max}\}$.
\end{enumerate}

Results of the ablation study are presented in Fig. \ref{fig:figA1}. Here, panels (a) and (b) report the MAE for exponent regression at $L=100$ and $L=500$, respectively, as a function of $a_{\max}$; panels (c) and (d) show the corresponding micro-F1 scores for diffusion-model classification. Obviously, for both trajectory lengths and both tasks, the performance curves for both the integer-only and the combined scale configurations exhibit a clear optimum around $a_{\max}=15$. Increasing $a_{\max}$ beyond this point does not provide systematic improvement and in some cases degrades the performance of wavelet representation, which indicates that very coarse scales contribute little additional useful information while potentially introducing redundancy or edge effects.

In addition, the comparison between integer-only and combined scales shows that augmenting the integer scales with the sub-unit scales $a\in \{0.1,0.2,\cdots,0.9\}$ consistently improves performance: across all tested $a_{\max}$ and both tasks, the combined-scale representation attains lower MAE and higher F1 than using only integer scales. Moreover, as highlighted by the stars in Fig. \ref{fig:figA1}, the sub-unit-only scale already achieves non-trivial performance levels in both MAE and F1, which demonstrates that the wavelet coefficients at $a<1$ indeed encode meaningful short-time structure.

Based on this ablation, we adopt the combined scale grid
$\{0.1,0.2,\ldots,0.9,1,2,\ldots,15\}$
as the default in all experiments reported in the main text. This choice balances coverage of short- and long-time behavior, avoids unnecessary proliferation of very coarse scales, and is empirically near-optimal for both exponent regression and model classification.

\begin{figure*}
\centering
\includegraphics[width=17.5cm]{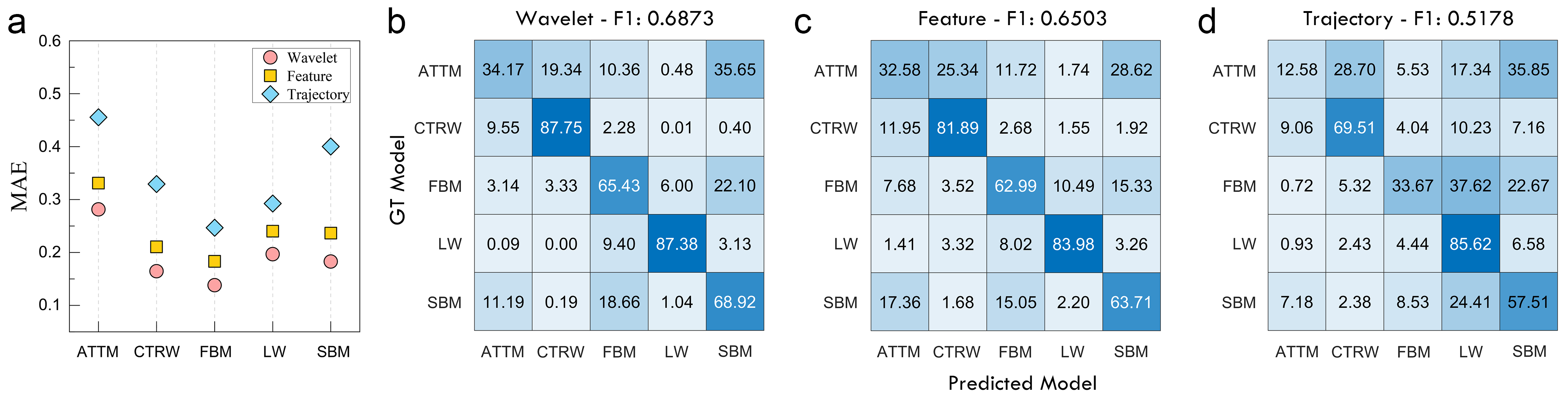}
\caption{Model-wise performance comparison on 2D simulated trajectories of length $L=100$ in the small-data regime ($N_{\rm train}=1000$). (a) MAE for diffusion-exponent regression across five specific models (ATTM, CTRW, FBM, LW, SBM). The wavelet representation consistently achieves the lowest error compared to feature-based and trajectory-based ones. (b)-(d) Confusion matrices for diffusion-model classification using wavelet-, feature-, and trajectory-based representations, respectively. Rows represent ground-truth (GT) labels, and columns represent predictions; values indicate the percentage of correct classifications (diagonal) and misclassifications (off-diagonal). The wavelet representation shows the strongest diagonal dominance and highest overall F1 score.}
\label{fig:figA3}
\end{figure*}

\section{Length distributions of experimental SPT trajectories}\label{SPTdata}

To document the heterogeneity in trajectory length in the experimental SPT dataset, Fig.~\ref{fig:figA2} summarizes the length distributions of trajectories recorded in F-actin networks with mesh sizes $\xi = 225, 250, 300, 550, 625,$ and $750~\mathrm{nm}$. Panels (a)-(c) correspond to the three smaller meshes ($225$, $250$, and $300~\mathrm{nm}$), where most trajectories contain more than $10^3$ time steps and some extend up to several thousand steps. Panels (d)-(f) show the distributions for the larger meshes ($550$, $625$, and $750~\mathrm{nm}$), where the trajectories are typically shorter, with lengths concentrated in the range of a few hundred time steps. These unbalanced length distributions motivate the fixed-length segmentation procedure described in Sec.~\ref{task2} when constructing datasets for supervised learning.

\section{Performance analysis on simulated trajectories across diffusion models}\label{simmodels}

To assess the universality of the wavelet representation, we further dissect the performance of the three representation families across five specific diffusion models in the andi-datasets library: ATTM, CTRW, FBM, LW, and SBM. The results for 2D simulated trajectories of length $L=100$ in the small-data regime ($N_{\rm train}=1000$) are summarized in Fig. \ref{fig:figA3}. For the diffusion-exponent regression task [Fig. \ref{fig:figA3}(a)], a consistent hierarchy is observed irrespective of the underlying physical mechanism: the wavelet representation achieves the lowest regression error for every model. Notably, the wavelet advantage is preserved across diverse dynamical regimes, ranging from non-ergodic processes (e.g., CTRW, ATTM) to ergodic Gaussian processes with long-range correlations (e.g., FBM), demonstrating that its efficacy is not tied to a specific type of anomalous diffusion.

For the diffusion-model classification task, the results for the three representations are shown as confusion matrices in Figs. \ref{fig:figA3}(b)-\ref{fig:figA3}(d). The wavelet representation [Fig. \ref{fig:figA3}(b)] exhibits the strongest diagonal dominance, yielding the highest overall micro-F1 score (0.6873). Misclassifications are largely confined to physically ambiguous pairs, such as the confusion between ATTM and SBM models, which is a known challenge for short trajectories due to their similar statistical signatures. The feature-based representation [Fig. \ref{fig:figA3}(c)] shows a qualitatively similar confusion pattern but with systematically lower diagonal accuracy (F1: 0.6503). In sharp contrast, the trajectory-based representation [Fig. \ref{fig:figA3}(d)] suffers from significant off-diagonal dispersion (F1: 0.5178). It struggles particularly with differentiating FBM from LW and SBM, as evidenced by the high misclassification rates in the corresponding rows.

Taken together, these model-specific analyses confirm that the superior performance of the wavelet representation is robust and universal. It effectively captures the distinct multiscale features of each diffusion mechanism, thereby outperforming feature- and trajectory-based baselines in both regression and classification tasks.

\section{Optimization of localization noise intensity for simulation-based models}\label{addnoise}

To reduce the distributional mismatch between the simulated training data and the experimental recordings, we explicitly add Gaussian localization noise into the simulation-based training pipeline. The noise intensity, parameterized by the standard deviation $\sigma_n$, is determined via a systematic grid search designed to minimize the MAE of the simulation-based models on the experimental test set. We construct a series of synthetic training sets for both the Sim-FBM and Sim-All configurations, each containing $10^5$ trajectories. For each configuration, we generate distinct versions by adding zero-mean Gaussian noise with standard deviations $\sigma_n \in \{0.00, 0.02, \ldots, 0.20\}$. We then train the WADNet model on each noise-augmented dataset and evaluate its diffusion-exponent regression performance (MAE) on the full experimental test set. This procedure is repeated for trajectory lengths $L=100$ and $L=200$ to ensure the robustness of the optimal parameter choice across different timescales.

\begin{figure}
\centering
\includegraphics[width=8.5cm]{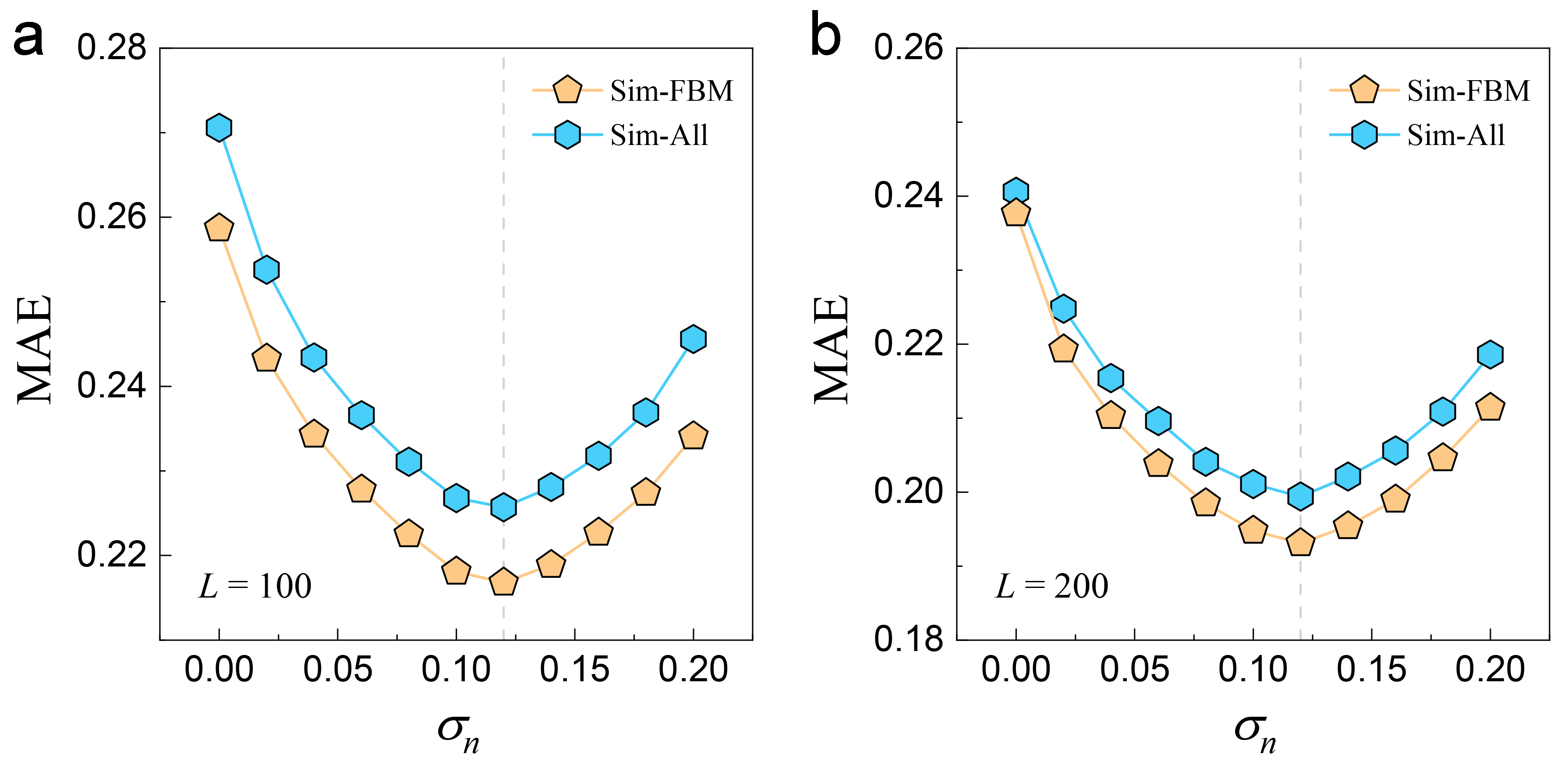}
\caption{MAE on the experimental test set with segments of length $L=100$ (a) and $L=200$ (b), as a function of the noise standard deviation $\sigma_n$. In both cases the MAE displays a U-shaped dependence on $\sigma_n$ with a consistent minimum at $\sigma_n=0.12$ (vertical dashed line).}
\label{fig:figA4}
\end{figure}

Fig. \ref{fig:figA4} summarizes the results of this optimization. For both $L=100$ [Fig. \ref{fig:figA4}(a)] and $L=200$ [Fig. \ref{fig:figA4}(b)], the MAE exhibits a clear U-shaped dependence on $\sigma_n$. In the absence of noise ($\sigma_n = 0$), the models suffer from large errors due to the lack of resilience against experimental imperfections. As noise is introduced, performance improves significantly, reaching a distinct minimum before degrading again as the noise overwhelms the signal. Crucially, for both the Sim-FBM and Sim-All models, and across both trajectory lengths, the optimal performance is consistently achieved at $\sigma_n=0.12$ (highlighted by dashed lines). Consequently, we fix this optimal noise level ($\sigma_n=0.12$) for all large-scale ($N_{\mathrm{train}}=10^{6}$) simulation-based training described in the main text.

\section{Ablation study on wavelet families and selection of representative scalograms}\label{waveletselection}

\begin{figure}
\centering
\includegraphics[width=7.2cm]{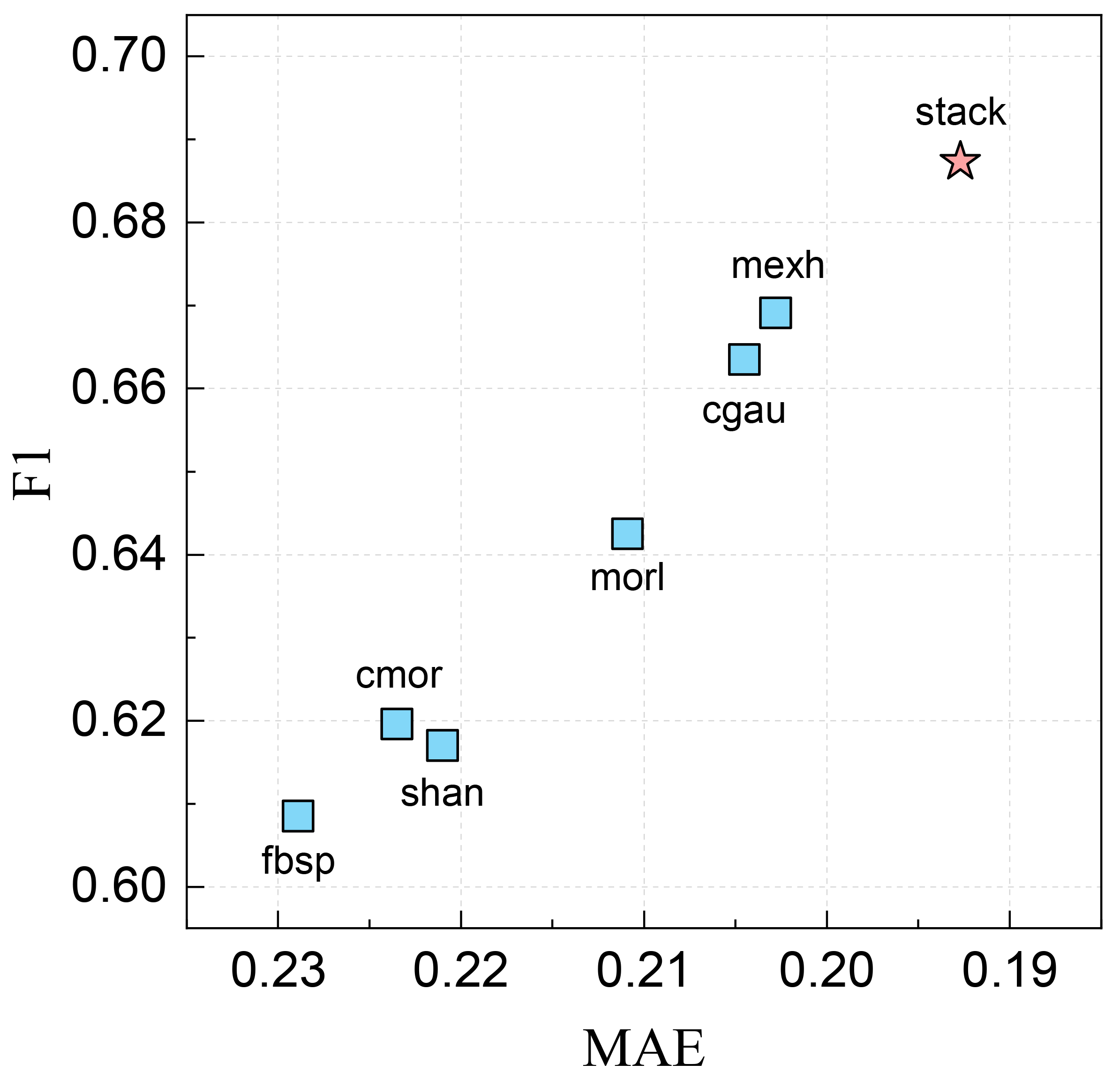}
\caption{Performance comparison between the full multi-channel wavelet representation (``stack'') and single-wavelet representations derived from individual wavelet families on the test set. The $x$-axis represents the MAE for diffusion-exponent regression, and the $y$-axis represents the micro-averaged F1 score for diffusion-model classification. The stacked representation achieves the best overall performance, demonstrating the benefit of ensemble learning from complementary wavelet bases.}
\label{fig:figA5}
\end{figure}

\begin{figure*}
\centering
\includegraphics[width=17.0cm]{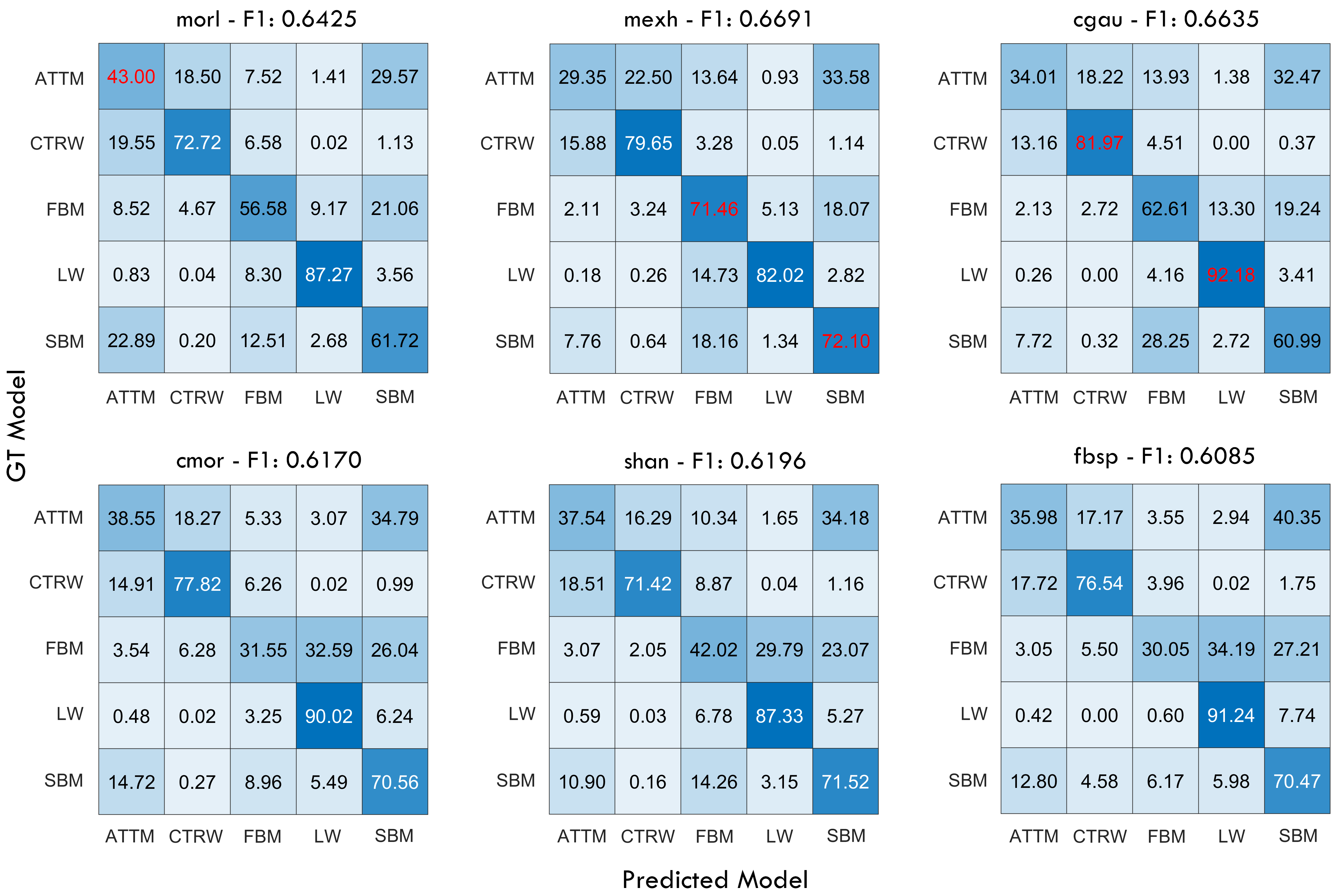}
\caption{Confusion matrices for diffusion-model classification using single-wavelet representations. Each panel corresponds to a model trained exclusively on scalograms from one wavelet family. The overall micro-F1 score is indicated in the title of each matrix. Red numbers highlight the highest diagonal accuracy achieved for each specific diffusion model across the different wavelet families. These maxima guide the selection of representative scalograms for visualization: morl is selected for ATTM; cgau for CTRW and LW; and mexh for FBM and SBM.}
\label{fig:figA6}
\end{figure*}

The wavelet representation proposed in this work is constructed by stacking the modulus scalograms from six distinct continuous wavelet families: real Morlet (morl), Mexican hat (mexh), complex Gaussian derivative (cgau), complex Morlet (cmor), Shannon (shan), and frequency B-spline (fbsp). The downstream vision models are always trained on this six-wavelet representation in order to exploit the complementary information provided by the different wavelets. For the interpretability analysis in Sec. \ref{interpretability}, however, we need to display individual scalograms when discussing scale fingerprints. This necessitates a criterion for selecting the most representative wavelet family for each diffusion model. In this appendix, we analyze the contribution of individual wavelet families to justify the multi-channel design and to guide the selection of these representative visualizations.

To assess the necessity of combining multiple wavelets, we first quantify the contribution of each wavelet family in isolation. Using 2D simulated trajectories with $N_{\rm train}=1000$ and $L=100$, we train separate EffNet models on single-wavelet representations, where the scalogram of only one mother wavelet is provided as input. Fig. \ref{fig:figA5} compares the test performance of these single-wavelet models with that of the full six-wavelet ``stack'' model. The stacked representation (marked by the star) consistently outperforms any single wavelet family in both diffusion-exponent regression (lowest MAE) and diffusion-model classification (highest F1), thereby validating the multi-channel stacking strategy employed in the main text.

For the purpose of visual interpretation, we then identify, for each diffusion model, the wavelet family that performs best when used alone. To this end, we analyze the class-wise performance using the confusion matrices of the single-wavelet classifiers, as shown in Fig. \ref{fig:figA6}. We adopt a best-in-class selection strategy: for each diffusion model, we select the mother wavelet that achieves the highest diagonal accuracy. The corresponding entries in Fig. \ref{fig:figA6} are highlighted in red. This strategy yields the following choices:

\begin{enumerate}
  \item ATTM: morl attains the highest diagonal accuracy $43.00\%$ for ATTM.
  \item CTRW and LW: cgau achieves the best performance for both CTRW and LW, with diagonal accuracies of $81.97\%$ and $92.18\%$, respectively.
  \item FBM and SBM: mexh gives the highest diagonal accuracy for FBM and SBM, with values of $71.46\%$ and $72.10\%$, respectively.
\end{enumerate}

These wavelet families are therefore used to generate the representative scalograms in Fig. \ref{fig:fig9} in Sec. \ref{interpretability}: morl for ATTM, cgau for CTRW and LW, and mexh for FBM and SBM. We stress that this choice is pragmatic and guided by the single-wavelet classification performance. The full six-wavelet representation remains superior for all quantitative tasks, but the selected wavelets provide clear and model-specific scale fingerprints that are well suited for qualitative interpretation.

\bibliography{Bibliography}

\end{document}